\documentstyle{article}

\newcommand{\thd}{\dot{\theta}}
\newcommand{\phz}{\phi_{\circ}}
\newcommand{\taut}{\tilde{\tau}}
\newcommand{\sct}{\tilde{S}_{C}}
\newcommand{\pal}{\large}
\newcommand{\palss}{\footnotesize }
\newcommand{\pals}{\normalsize}
\newcommand{\palb}{\normalsize\bf}
\newcommand{\palbb}{\large\bf}

\begin{document}

\pals

\hfill PUPT-1435
\hfill cond-mat/9401072

\bigskip\bigskip\bigskip

\leftline{\palbb S{\palb TATISTICAL} M{\palb ECHANICS AND}
V{\palb ISUAL} S{\palb IGNAL} P{\palb ROCESSING}}
\bigskip\bigskip
\leftline{M{\palss ARC} P{\palss OTTERS}\raisebox{.6 ex}{\scriptsize
(1,2)} {\palss AND }W{\palss ILLIAM}
B{\palss IALEK}\raisebox{.6 ex}{\scriptsize (1)}}
\bigskip\bigskip
\leftline{\raisebox{.6 ex}{\scriptsize (1)}
NEC Research Institute, 4 Independence Way, Princeton, New Jersey 08540}
\bigskip
\leftline{\raisebox{.6 ex}{\scriptsize (2)}
Department of Physics, Princeton University, Princeton, New Jersey 08544}

\bigskip\bigskip
\leftline{January 1994}
\vfill
\bigskip
\hrule
\bigskip\bigskip

The nervous system solves a wide variety of problems in signal
processing.  In many cases the performance of the nervous system is so
good that it apporaches fundamental physical limits, such as the limits
imposed by diffraction and photon shot noise in vision.
In this paper we show how to use the language of statistical field theory
to address and solve problems in signal processing, that is problems in
which one must estimate some aspect of the environment from the data
in an array of sensors.  In the field theory formulation the optimal
estimator can be written as an expectation value in an ensemble where
the input data act as external field.  Problems at low signal-to-noise
ratio can be solved in perturbation theory, while high signal-to-noise
ratios are treated with a saddle-point approximation.
These ideas are illustrated in
detail by an example of visual motion estimation which is chosen to
model a problem solved by the fly's brain.  In this problem the optimal
estimator has a rich structure, adapting to various parameters of the
environment such as the mean-square contrast and the correlation time of
contrast fluctuations.  This structure is in qualitative accord with
existing measurements on motion sensitive neurons in the fly's brain,
and we argue that the adaptive properties of the optimal estimator may
help resolve conlficts among different interpretations of these data.
Finally we propose some crucial direct tests of the adaptive behavior.

\newpage
\section{Introduction}

Imagine walking along a busy city street.  As we walk, we are almost
unaware of the myriad tasks which our brains are performing: We use a
combination of visual and vestibular signals to keep ourselves upright,
sensors in our feet and leg muscles help adjust our stride to the terrain,
we listen for cars and people behind us, vision helps us recognize the
caf\'e we are approaching and identifies our friend who will meet us there,
and all these senses combine to provide us with a trajectory which reaches
our goal and avoids obstacles.  All of these tasks involve {\it signal
processing,} and we have the qualitative impression that we (and other
animals) are quite good at solving these problems.  The goal of this paper
is to show that statistical mechanics provides the natural language for
formulating and solving signal processing problems, and that the structure
of the statistical mechanics models provides a predictive theory of how
real brains solve the corresponding problems.

\subsection{Physical limits and biological signal processing}

Signal processing is in essence a physics problem.  As an example, in
vision the precise formulation of any task must begin with the fact that
images are blurred by diffraction and corrupted by photon shot noise.
These irreducible limitations in the quality of the input signal limit the
reliability with which the brain (or any device) can estimate what is
really going on in the visual environment---we can never be truly certain
of what we are looking at, nor can we know its exact trajectory, and so on.
 There are physical limitations from the hardware as well---cells of a
certain size are bound to generate electrical noise, signals are passed
from one point to another along cables of limited information capacity, ...
 .  Remarkably, these physical limitations to the reliability of
computation are actually relevant to the operation of real brains.  Indeed
there are many cases where the performance of the nervous system is
essentially equal to the limit that one calculates from first principles
\cite{annrevs,bialek-plb}; examples range from photon counting in toads and
humans \cite{barlow-ferrier,donner-review} to visual motion estimation in
flies \cite{science,stepdisc-nips} to acoustic coding in frogs and crickets
\cite{europhy} and echo delay estimation in bats \cite{simmons-nano}.
These observations strongly suggest that a theory of optimal signal
processing will help us understand the computational strategies of brains.

\subsection{Choosing a problem}

Of the many examples of signal processing in the nervous system, our
discussion is motivated most directly by the problem of motion estimation
in the fly's visual system.  In many insects the visual system produces
movement signals which are used to control the flight muscles and thereby
stabilize flight.  In the case of flies, visually guided flight behavior
has been studied both by measuring flight paths during natural behaviors
\cite{land+collett,wagner1,wagner2,wagner3} and by examining the torques
produced by flies hanging from a torsion balance in response to movements
of controlled patterns across the visual field
\cite{heisenberg-wolf,quartrevI}.  The input to the motion computation
comes from a single class of photoreceptor cells which are arrayed in the
regular lattice of the compound eye, and the signal and noise properties of
these cells are extremely well characterized.  The output of the movement
computation can be monitored in a handful of identified cells of the lobula
complex \cite{franceschini-facets,hausen84}, and destruction of individual
lobula plate neurons produces clear deficits in the fly's opto-motor
behavior \cite{hausen+wehrhahn}.  The fact that these neurons are
``identified'' has a technical meaning: cells of essentially identical
morphology occur in each fly, and these cells have responses to visual
stimuli which are quantitatively reproducible from individual to
individual.  Thus the cells can be named and numbered based on either
structure or function; under favorable conditions one can record from the
cell H1, which codes wide field, horizontal movements of the visual field,
for periods of up to five days \cite{rob-thesis}.  The accessibility of
quantitative measurements at each of several layers in the nervous
system---photoreceptors, second order neurons, motion-sensitive cells,
flight behavior---makes the fly a nearly ideal testing ground for theories
of neural signal processing.

\subsection{Relation to previous work}

Signal processing is of course an enormous field with a diverse set of
applications \cite{leshouches85,papoulis,vantrees}.  At the core of this
field is the description of random functions, whether they be functions of
time (e.g., sounds) or space (e.g., images).  Nonetheless, the bulk of the
literature on signal processing does not seem to make extensive use of the
functional integral methods which seem so natural from the perspective of
statistical mechanics.  In 1988, however, Zweig and Lackner
\cite{lackner+zweig} studied a model signal processing problem,
reconstructing the configuration of a randomly moving plate using a set of
sparse and noisy observations of displacements at particular points along
the plate.  By choosing the random motions from the Boltzmann distribution
they emphasized the connection to statistical mechanics.

Traditional approaches to the problem of ``optimal estimation'' begin by
defining a class of estimation strategies and then use variational methods
to search this class for the best estimator.  This approach goes back to
Wiener and Kolmogorov, who found the optimal linear filters for recovering
and extrapolating signals in noisy backgrounds.  In this analysis the
causality of filters is a crucial constraint, and hence the key
mathematical ingredient is the role of the causal analytic structure of
filters in solving the integral equations which result from the variational
definition of optimality.  The Wiener-Kolmogorov results essentially close a
broad class of questions related to optimal estimation by linear filters,
but leave completely open problems where the optimal estimator may be a
non-linear function of the input data.  In fact, the classical approach
gives us little hint about the conditions for the importance of
non-linearity.  We shall see that the statistical mechanics approach gives
us tools for going beyond linear (or lowest order non-linear) estimators.
Rather than searching through a limited class of estimators for a local
optimum, the mapping to a statistical mechanics problem allows us to
directly construct the globally optimal estimator.

In the case of vision, it is widely recognized that unconstrained
variational approaches to estimation result in ill-posed problems
\cite{koch+al}.  The ``regularization'' of these ill-posed problems can be
given a probabilistic interpretation, in which the optimal estimator
reflects a compromise between fitting the available data and taking account
of {\it a priori} knowledge about the expected distribution of incoming
signals; this is the same structure described by Zweig and Lackner in their
model problem.  Different regularization terms thus represent different
hypotheses about the statistical structure of the visual environment, but
there has been very little experimental work to characterize these
statistics \cite{scaling}.  Following Geman and Geman \cite{geman+geman},
there has been considerable focus on ``Markov random field'' models for
image statistics, which are known to be equivalent to Boltzmann
distributions with local Hamiltonians, but there has been relatively little
work exploiting the full statistical mechanics structure of even these
simple models.  In particular, if one takes the statistical models
seriously as models of the problems that are solved by real brains, there
is the clear prediction that the signal processing strategies of the brain
must adapt to the statistical structure of the environment.  This sort of
adaptation is much richer than that conventionally described in the
biological literature, and we shall explore these predicted effects in some
detail since we feel that they constitute the key experimental signature of
optimal estimation.

The connection between statistical mechanics and signal processing has been
used in previous work to study vision at very low photon flux, where the
low signal-to-noise ratio (SNR) leads to a universal ``pre-processing''
stage in any estimation task \cite{bipolar-bj,bipolar-nips,fred-thesis}.
The structure of this pre-processing step is in good agreement with the
responses of cells in the first stage of retinal signal processing, and as
far as we know this is the first example of a successful parameter-free
prediction of neural responses.  Most sensory systems encode incoming
signals in sequences of discrete pulses, termed spikes or action
potentials, and the problem of decoding these pulse sequences can again be
given a statistical mechanics formulation \cite{bialek+zee}; the resulting
predictions concerning the algorithms for optimal decoding have been
confirmed in experiments on a wide variety of organisms \cite{bits+brains}.
Finally, the low SNR limit of motion estimation in fly vision has also been
studied \cite{santafe,bialek-plb,optfly}, and this analysis was crucial in
establishing that the fly does make optimal estimates, but we shall see
that by restricting attention to low SNR one misses a large amount of
structure which is likely relevant under natural conditions.  Attempts to
go beyond the low SNR perturbation theory have been confined to model
problems \cite{bialek+zee2,beyond-nyquist}.

\section{Simple Examples of Signal Processing}
\label{s:simple}
A typical signal processing problem consists of extracting information from
the output of a device corrupted by noise.  The signal might be encoded in
the data in a complex way or might even be present only statistically (as
in the case of prediction of time series).  The task is then, given the
statistical properties of the quantities of interest, to find the optimal
estimate of the signal using the data.  To do so, one must decide on
a definition of the quality of an estimate.  There are two obvious choices:
maximum likelihood and least-mean-square error.  The maximum likelihood
estimate corresponds to the signal that has the highest probability given
the observation of the data.  The least-mean-square estimator is the one
that would on average minimize $\chi^2$, the distance square to the true
value; it is equal to the conditional mean.  There are other cost
functions whose minimizations lead to different estimators.  Before we
proceed with our main problem, let us explain our approach on a few simple
models where everything can be computed exactly.

\subsection{One variable estimation}

Consider the following problem: estimate a signal
consisting  of a single number $s$
using the output $y$ of a detector contaminated by noise, $y=s+\eta$. Since
the noise is random and the true signal is unknown to the observer,
this problem is a probabilistic one.
To find the best estimate of the signal,
we need to construct the probability of $s$
given the output $y$. We use Bayes' theorem to write down the conditional
probability of $s$ given $y$:
\begin{equation}
P[s|y]=\frac{P[y|s]P[s]}{P[y]}. \label{eq:bayes}
\end{equation}

$P[y|s]$ describes the detection process: how the output is related to the
input signal and the probabilistic nature of the noise.  $P[s]$ reflects
the {\em a-priori} knowledge of the signal, and $P[y]$ is independent of
$s$ and thus serves as a normalization.

Suppose first that both signal and noise are chosen
from independent Gaussian distributions with variance $S$ and $N$
respectively, i.e.
\begin{equation}
P[s]=\frac{1}{\sqrt{2\pi S}}\exp\left[-\frac{s^2}{2S}\right] \mbox{ and }
P[y|s]=\frac{1}{\sqrt{2\pi N}}\exp\left[-\frac{(y-s)^2}{2N}\right].
\end{equation}
After a little bit of algebra, we find
\begin{equation}
P[s|y]=\sqrt{\frac{S+N}{2 \pi NS}}
\exp\left[-\frac{(S+N)(s-\frac{Sy}{S+N})^2}{2NS}\right].
\label{eq:toyps}
\end{equation}
The conditional probability distribution is thus also Gaussian with mean
 $\frac{Sy}{S+N}$ and variance $\frac{NS}{S+N}$. In this example,
the maximum and the average of the conditional probability distribution
are equal and are given by
\begin{equation}
s_e=\left(\frac{S}{S+N}\right)y.
\label{eq:toy-est}
\end{equation}
The best estimate of the input is thus a linear function of the detector
output. This will be true in general whenever the probability distributions
are Gaussian. It is not necessarily the case when either the signal or the
noise is non-Gaussian or when the output is a non-linear function of the
signal.

To illustrate the emergence of non-linear estimators, consider a
small modification of our toy-model. Suppose now that the signal is
positive and taken from an exponential distribution with mean $1/ \alpha$
and that the noise is Gaussian and additive but has a variance proportional
to the signal (i.e. $N=\beta s$).The analog of Eq. (\ref{eq:toyps}) is now
\begin{equation}
P[s|y]=\frac{1}{Z\sqrt{s}}
\exp\left(-\alpha s-\frac{(y-s)^2}{2\beta s}\right),
\label{eq:nongauss}
\end{equation}
where $Z$ is an $s$-independent normalization factor.
The maximum likelihood estimator, which maximizes Eq.
(\ref{eq:nongauss}), is given by
\begin{equation}
s_{e}^{\mbox{\scriptsize ML}}=
\frac{
 \sqrt{\beta^2+4y^2(1+2\alpha\beta)}-\beta
}{
 2(1+2\alpha\beta)
}.
\end{equation}

To compute the least mean-square estimator, we need to average $s$ with
respect to $P[s|y]$.  After some algebra and a trip to our favorite
integral table we find
\begin{equation}
s_{e}^{\mbox{\scriptsize LMS}}=\frac{\beta}{1+2\alpha\beta}
+\frac{|y|}{\sqrt{1+2\alpha\beta}}.
\end{equation}

In this example, the two definitions of optimal estimation lead to
different results; both are non-linear functions of the output variable
$y$.  This of course comes from the non-Gaussian nature of the signal and
the dependence of the noise upon the signal. Amusingly, in this case the
optimal estimator is not even monotonic in the detector output $y$.

\subsection{Causal linear filtering}
\label{ss:causal}

Let us now consider the estimation of continuous signals by doing the time
dependent version of our Gaussian toy-model.  Suppose now that $s(t)$ and
$\eta(t)$ are functions of time chosen from Gaussian distributions with
power spectra $S(\omega)$ and $N(\omega)$. In functional integral
language \cite{feynman+hibbs,kac}:
\begin{equation}
P[s]=\frac{1}{Z_S} \exp\left[ - \frac{1}{2}
\int_{-\infty}^{\infty}
\frac{d\omega}{2 \pi} \frac{\left| s(\omega) \right|^{2}}{
S(\omega)} \right],
\end{equation}
where $Z_S$ is an ill-defined normalization constant which as usual
never enters in the computations.  There is a similar equation for
$P[y|s]$.  We find the conditional probability distribution
\begin{equation}
P[s|y]  =\frac{1}{Z} \exp \left[ - \frac{1}{2} \int_{-\infty}^{\infty}
\frac{d\omega}{2 \pi} \frac{\left| y(\omega) - s(\omega)
\right|^{2}}{N(\omega)} - \frac{1}{2}
\int_{-\infty}^{\infty}
\frac{d\omega}{2 \pi} \frac{\left| s(\omega) \right|^{2}}{
S(\omega)} \right].
\end{equation}
To find $\langle s(\omega) \rangle_{y}$ we compute the
``equation of motion'' and find
\begin{equation}
s_{e}(\omega) = \left(\frac{S(\omega)}{S(\omega) + N(\omega)}\right)
y(\omega),
\end{equation}
which can be written in the time domain:
\begin{equation}
s_{e}(t)=\int_{-\infty}^{\infty}d\tau g(\tau) y(t-\tau) \mbox{  where  }
g(\tau)=\int_{-\infty}^{\infty}
\frac{d\omega}{2
\pi} e^{-i \omega \tau}  \frac{S(\omega)}{S(\omega) +
N(\omega)}.
\label{eq:acausal}
\end{equation}
We see that the problem of estimating continuous functions of time is just
the problem of estimating the independent Fourier components; for each
component we have an equation identical to Eq.  (\ref{eq:toy-est}) for the
estimation of a single variable.  It should be clear that this simplicity
is tied to the assumption of Gaussian distributions for both the signal and
the noise.

The filter $g(\tau)$ is acausal, since it extends symmetrically both in the
past and in the future.  If we actually want to build a device which takes
the $y(t)$ as input and returns an estimate of the signal $s(t)$, then this
device must be causal.  It is natural to ask what is the optimal causal
estimator, one that would required knowledge of the output $y(t)$ only up
to the present or, more generally, up to a small delay time $\tau_\circ$ in
the future.  The delay time $\tau_\circ$ could be negative, that would be a
predictor instead of an estimator, but the same method would apply.  This
problem was solved by Wiener \cite{wiener-time} and Kolmogorov
\cite{kolmogorov,kolmogorov2}.  We give here another derivation in a
slightly different context.  Wiener assumed linear filtering and found that
the optimal causal filter only required knowledge of the signal and noise
power spectra.  We will assume that the signal and noise are chosen from
Gaussian distribution with specified power spectra and find that the
optimal estimator is a linear filter.

The task is to estimate $s(-\tau_\circ)$ given the output up to time $t=0$.
 Once again we write our estimator as the conditional average of the random
variable $s(-\tau_\circ)$ upon observing $y(t<0)$.  This is equivalent to
averaging first with respect to the conditional probability of $s$ given
all of $y$ and then integrating out $y^+(t)$ using the conditional
probability $P[y^+(t)|y^-(t)]$ where
\begin{equation}
\begin{array}{rclrcl}
y^{+}(t>0)&=& y(t)& y^{-}(t>0)&=&0\\
y^{+}(t<0)&=&0 & y^{-}(t<0)&=&y(t).
\end{array}
\end{equation}

We need to write down the probability distribution of
$y^{+}(t)$ given $y^{-}(t)$.
\begin{equation}
P[y^{+}(t)|y^{-}(t)]=\frac{P[y(t)]}{P[y^{-}(t)]}
\end{equation}
\begin{equation}
P[y(t)]=\frac{1}{Z}\exp \left[ -\frac{1}{2}  \int_{-\infty}^{\infty}
\frac{d\omega}{2 \pi} \frac{\left| y(\omega)  \right|^{2}}{
\left(N(\omega)+S(\omega)\right)}\right]
\end{equation}
The key trick \cite{wiener-time} is to write
\begin{equation}
\left| \Psi(\omega)\right|^{2}=\frac{1}{N(\omega)+S(\omega)},
\end{equation}
where $\Psi(\omega)$ is chosen such that it has no poles in
the upper
half plane. The probability distribution in the time domain
is then
\begin{equation}
P[y(t)]=\frac{1}{Z}\exp \left[ -\frac{1}{2}
\int_{-\infty}^{\infty}d\tau
\left|\int_{-\infty}^{\infty}\frac{d\omega}{2\pi}
y(\omega)\Psi(\omega) e^{-i \omega \tau}  \right|^{2}\right].
\end{equation}

The function $x(\tau)$ defined by
\begin{equation}
x(\tau)=\int_{-\infty}^{\infty}\frac{d\omega}{2\pi}
y(\omega)\Psi(\omega) e^{-i \omega \tau}
\end{equation}
is causal (i.e. it only depends on the values of $y(t)$ for
$t<\tau$).
\begin{eqnarray}
P[y^{-},y^{+}]=\frac{1}{Z}\exp \left[ -\frac{1}{2}
\int_{-\infty}^{0}d\tau
\left|\int_{-\infty}^{\infty}\frac{d\omega}{2\pi} y^{-
}(\omega)\Psi(\omega) e^{-i \omega \tau}
\right|^{2}\right.\nonumber \\
\left.\mbox{}-\frac{1}{2}  \int_{0}^{\infty}d\tau \left|\int_{-
\infty}^{\infty}\frac{d\omega}{2\pi}(y^{-
}(\omega)+y^{+}(\omega))\Psi(\omega) e^{-i \omega \tau}
\right|^{2}\right]
\end{eqnarray}
\begin{equation}
P[y^{-}]=\int [dy^{+}] P[y^{-},y^{+}]
\end{equation}

To do the integral over $y^{+}(t)$ we change variables to $x^{+}(\tau)$
defined as $x(\tau)$ for $\tau>0$.  We notice that this is a linear
transformation, and since $x(\tau)$ is causal the transformation matrix is
lower triangular.  Therefore the Jacobian of this transformation is
independent of $y^{-}(t)$.  The Gaussian integral over $x(\tau)$ gives an
overall factor that can be absorbed in the normalization of our probability
distribution.
\begin{equation}
P[y^{-}]=\frac{1}{Z}\exp \left[ -\frac{1}{2}  \int_{-\infty}^{0}d\tau
\left|\int_{-\infty}^{\infty}\frac{d\omega}{2\pi} y^{-
}(\omega)\Psi(\omega) e^{-i \omega \tau}  \right|^{2}\right]
\end{equation}
\begin{equation}
P[y^{+}(t)|y^{-}(t)]=\frac{1}{Z}\exp \left[ -\frac{1}{2}
\int_{0}^{\infty}d\tau
\left|\int_{-\infty}^{\infty}\frac{d\omega}{2\pi} \left(y^{-}
(\omega)+y^{+}(\omega)\right)\Psi(\omega) e^{-i \omega \tau}
\right|^{2}\right]
\label{eq:py+}
\end{equation}
As mentioned earlier, we first average $s$ using $P[s|y]$ and then average the
result with respect to $P[y^+|y^-]$. The result of the first averaging is the
acausal estimator (\ref{eq:acausal}) evaluated at $t=-\tau_\circ$,
\begin{equation}
s_{\mbox{\scriptsize nc}}(-\tau_\circ) = \int_{-\infty}^{\infty}
\frac{d\omega}{2
\pi} e^{i \omega \tau_\circ}  y(\omega)
S(\omega)\Psi(\omega)\bar{\Psi}(\omega).
\label{eq:snc2}
\end{equation}
Since $s_{\mbox{\scriptsize nc}}(-\tau_\circ)$ is linear in $y(t)$,
averaging with respect to (\ref{eq:py+}) amounts to replacing $y(t)$ by its
average value which solves the {\em equation of motion}\/:
\begin{equation}
\int_{-\infty}^{\infty}\frac{d\omega}{2\pi} \left(y^{-}
(\omega)+y^{+}(\omega)\right)\Psi(\omega) e^{-i \omega \tau} = 0
\mbox{ for } \tau>0.
\label{eq:eqofmotion}
\end{equation}
Imposing (\ref{eq:eqofmotion}) amounts to replacing $y(\omega)\Psi(\omega)$ by
\begin{equation}
\int_{-\infty}^{0}d\tau e^{i\omega\tau}\int_{-
\infty}^{\infty}\frac{d\omega^{\prime}}{2\pi} y
(\omega^{\prime})\Psi(\omega^{\prime}) e^{-i \omega^{\prime} \tau}.
\end{equation}
in equation~(\ref{eq:snc2}). The causal estimator is therefore
given by
\begin{equation}
s_{\mbox{\scriptsize c}}(-\tau_\circ) = \int_{-\infty}^{\infty}
\frac{d\omega}{2
\pi} e^{i \omega \tau_\circ}
S(\omega)\bar{\Psi}(\omega)\int_{-\infty}^{0}d\tau e^{i\omega\tau}\int_{-
\infty}^{\infty}\frac{d\omega^{\prime}}{2\pi} y
(\omega^{\prime})\Psi(\omega^{\prime}) e^{-i \omega^{\prime} \tau}
\end{equation}

We can easily extend this to the slightly more general case where the
output if known up to time $t$ and we want to estimate the signal at time
$t-\tau_\circ$.  After rewriting things a little bit, we obtain
\begin{equation}
s_{\mbox{\scriptsize c}}(t-\tau_\circ) = \int_{-\infty}^{\infty}
\frac{d\omega}{2\pi} e^{-i \omega t}y(\omega)k(\omega).
\end{equation}
\begin{equation}
k(\omega)=\Psi(\omega)\int_{0}^{\infty}d\tau
e^{i\omega\tau}\int_{-\infty}^{\infty}\frac{d\omega^{\prime}}{2\pi}
e^{i\omega^{\prime}
(\tau_\circ-\tau)}
S(\omega^{\prime})\bar{\Psi}(\omega^{\prime})
\end{equation}
We recognize this result as the Wiener filter \cite[p. 86]{wiener-time}.
Note that we are using a different definition
of $\Psi(\omega)$ and the opposite sign convention in Fourier transforms.
But more importantly, our derivation shows that non-linear terms would
not help the causal estimation of Gaussian signals with Gaussian noise.

\section{Optimal Rigid Motion Estimation}
\label{s:Motion}

We now turn to the problem of visual motion estimation. Although
our primary motivation is to present a theory of the problem solved by
the fly's  visual system, our formulation is fairly general and should be
applicable to any visual system, or even to the design of artificial
systems. It is important to realize that our ``model'' is a model of the
problem the fly is solving, not a model of the neural circuitry which
implements the solution. Thus we need not concern ourselves with the
details of fly neuroanatomy; what is important is that we capture the
essential features of the fly's environment which make the problem of
motion estimation non-trivial.

\subsection{The model}

The fly wants to estimate its own rotation as it tries to fly in a
straight line but is pushed around by the wind.  If the fly flew in a
straight line it would see a contrast pattern $C[x,t]$ that changes both as
a function of time $t$ and position $x$ in the visual field; we define
contrast as $ C =(I-I_{o})/I_{o}$ where $I$ is the light intensity at a
point and $I_{o}$ its average value.  Since we are interested only in
horizontal motion estimation we give a one-dimensional description, where
$x$ is just the azimuthal angle or yaw.  When the fly turns along some
angular trajectory $\theta(t)$ it sees a modified contrast pattern
$C[x-\theta(t),t]$ then each photoreceptor cell produces a voltage given by
\begin{equation}
V_{n}(t)=\int T(\tau)\int M(x_{n}-x)C[x-\theta(t-\tau),t-\tau]
+\delta V_{n}(t),
\label{eq:photo}
\end{equation}
where $T(\tau)$ is the temporal impulse response of the cell, $M(x)$ the
angular acceptance profile of the photoreceptor and $\delta V_{n}(t)$ is
the voltage noise.  The photoreceptors form an array of size $N$ and of
angular separation $\phz$.  We will assume that the noise is independent in
each photoreceptor, Gaussian and additive with power spectrum $N(\omega)$.
As in the previous section, we use Eq.  (\ref{eq:bayes}) to write the
conditional probability distribution where now $\theta(t)$ is the signal
and $\{V_n(t)\}$ are the detector output.  Equation (\ref{eq:photo}) and
the voltage noise power spectrum determine the probability of observing
$\{V_n(t)\}$ given a trajectory $\theta(t)$ and a contrast pattern
$C(x,t)$.  Since this contrast pattern is unknown to the observer, one
needs to average over all possible such contrast.
\begin{equation}
P[\theta(t)|\{V_n(t)\}]=\frac{1}{Z}\left\langle\exp\left[{-}\sum_n\int
\frac{d\omega}{2\pi}\frac{\left|V_n(\omega)-\bar{V}_n(\omega)\right|^2}
{2N(\omega)}\right]\right\rangle_C P\left[\theta(t)\right],
\label{eq:PthgV}
\end{equation}
where
\begin{equation}
\bar{V}_n(\omega)=T(\omega)\int dt e^{i\omega t}
\int dx M(x-x_n)C(x-\theta(t),t).
\end{equation}
Brackets mean the average over $P[C(x,t)]$, the spatio-temporal contrast
distribution.  For now we will keep the distribution $P[C(x,t)]$ and
$P[\theta(t)]$ as general as possible, except that we will assume
$P[\theta(t)]$ to be left-right and time reversal invariant.  We will
choose the optimal estimator to be the average $\thd(t)$ in the
distribution (\ref{eq:PthgV}) which, as mentioned in the first section,
minimizes $\chi^2$.
We can expand the norm square in (\ref{eq:PthgV})
and drop the $|V_n|^2$ term since it does not depend on $C(x,t)$ nor
$\theta(t)$ and can be absorbed in the overall normalization,
\begin{equation}
\thd_e(t)=\frac{1}{Z(V)}\left\langle\thd(t)\exp\left[{-}\sum_n\int
\frac{d\omega}{2\pi}\frac{\left|\bar{V}_n(\omega)\right|^2}
{2N(\omega)}-\frac{\bar{V}_n^*(\omega)V_n(\omega)}
{N(\omega)}\right]\right\rangle_{C,\theta},
\label{eq:thde}
\end{equation}
where
\begin{equation}
Z(V)=\left\langle\exp\left[{-}\sum_n\int
\frac{d\omega}{2\pi}\frac{\left|\bar{V}_n(\omega)\right|^2}
{2N(\omega)}-\frac{\bar{V}_n^*(\omega)V_n(\omega)}
{N(\omega)}\right]\right\rangle_{C,\theta}.
\end{equation}
The first term in the exponential modifies the a-priori independent action
for $C$ and $\theta$ by introducing a coupling between the two, while the
second couples them to an external field $V_n(\omega)$.  In this model, the
optimal estimator is the expectation value of an operator within a
statistical theory of two coupled fields (velocity and contrast) in the
presence of an external field (photoreceptor voltages).  This appears to be
the general structure of signal processing problems from the statistical
mechanics point of view---input data act as external fields, and optimal
estimators are expectation values.  Thus the construction of optimal
estimators is exactly the problem of constructing response functions.

\subsection{Perturbation theory}
\label{ss:perturbation}

As always in field theory it is not an easy task to write down an exact
expression for Eq.  (\ref{eq:thde}).  But fortunately there exists a
potentially small parameter in which one can do perturbation theory, namely
the signal-to-noise ratio (SNR).  If the mean receptor voltages
$\bar{V}_n(\omega)$ are small compared to the typical voltage noise
($\propto \sqrt{N(\omega)}$) then (following Ref.  \cite{optfly})we can
expand the exponential in (\ref{eq:thde}) and write the estimator as a
perturbation series in $\bar{V}_n$:
\begin{eqnarray}
\thd_e(t)&=&\sum_n\int\frac{d\omega}{2\pi}V_n(\omega)\frac{\langle\thd(t)
\bar{V}_n^*(\omega)\rangle}{N(\omega)} \nonumber \\
& & \mbox{}+\frac{1}{2}\sum_{nm}\int\frac{d\omega}{2\pi}
\int\frac{d\omega'}{2\pi}
V_n(\omega)V_m(\omega')\frac{\langle\thd(t)
\bar{V}_n^*(\omega)\bar{V}_m^*(\omega')\rangle}{N(\omega)N(\omega')}
\nonumber \\
& & \mbox{}-\frac{1}{2}\sum_{n}\int\frac{d\omega}{2\pi}\frac{\langle\thd(t)
|\bar{V}_n^*(\omega)|^2\rangle}{N(\omega)}+\ldots
\label{eq:perturbation}
\end{eqnarray}
The first term vanishes because it is linear in contrast and the average
contrast is zero by definition.  The last term is also zero because it is
odd under time reversal symmetry.  To compute the second term it is useful
to introduce a spatial Fourier representation; we find
\begin{eqnarray}
\langle\thd(t)\bar{V}_n^*(\omega)\bar{V}_m^*(\omega')\rangle
&=&T^*(\omega)T^*(\omega')
\int\frac{dk}{2\pi}|M(k)|^2e^{ik(x_n-x_m)} \nonumber \\
&&\qquad\times  \int d\tau\int d\tau' S_C(k,\tau-\tau')
e^{i(\omega\tau+\omega'\tau')} \nonumber\\
&&\qquad\times\langle\thd(t)e^{-ik[\theta(\tau)-\theta(\tau')]}\rangle .
\label{eq:thdVV}
\end{eqnarray}
where we defined the contrast two-point function $S_C(k,t)$ as
\begin{equation}
\langle C(k,t)C(k',t')\rangle=2\pi\delta(k+k')S_C(k,t-t'),
\end{equation}
assuming spatial and temporal translation invariance.  Actually, spatial
translation invariance might not always hold.  In the case of small angular
jitter on top of forward motion, there is a special point in the visual
field, namely the direction of the forward motion.  Relaxing this
assumption amounts to having $S_C$ depend on both $k$ and $k'$ which would
result in having two $k$ integrals in (\ref{eq:thdVV}).  This would not
change significantly the conclusions of this section, but might obscure
even more the formulae.  We can't say much about the function
$F(k,t,\tau,\tau')$ defined by
\begin{equation}
F(k,t,\tau,\tau')=\langle\thd(t)e^{-ik[\theta(\tau)-\theta(\tau')]}\rangle
,
\end{equation}
without being more specific about $P[\theta(t)]$. We will consider concrete
examples in the next section. Note however that the form of $F$ depends only
on the statistics of the trajectories and that
under our assumption of parity invariance, it must be antisymmetric in $k$.
The optimal velocity estimator, to lowest order in SNR is therefore given by
\begin{equation}
\thd_e(t)=\sum_{nm}\int\frac{d\omega}{2\pi}\int\frac{d\omega'}{2\pi}
g_{nm}(\omega,\omega')V_n(\omega)V_m(\omega'),
\label{eq:est-lowsnr}
\end{equation}
\begin{eqnarray}
g_{nm}(\omega,\omega')&=&\frac{T^*(\omega)T^*(\omega')}{N(\omega)N(\omega')}
\int\frac{dk}{2\pi}|M(k)|^2e^{ik(x_n-x_m)} \nonumber \\
&&\quad\times  \int d\tau\int d\tau'
e^{i(\omega\tau+\omega'\tau')}S_C(k,\tau-\tau')F(k,t,\tau,\tau').
\end{eqnarray}
Poggio and Reichardt pointed out \cite{quartrevII,quartrevI} that a
velocity estimator, as any other functional, could be written as Volterra
series in the photoreceptor output.  They also argued that the simplest
term that would give the proper directional sensitivity is an antisymmetric
correlator.  Here we have found that the dominant term of the optimal
velocity estimator at low SNR is an antisymmetric correlator whose temporal
and spatial characteristics are tuned to the statistics of trajectories,
images (spatio-temporal contrast patterns) and photoreceptor noise.  The
only statistical feature of real moving images needed to compute this
estimator is their spatio-temporal correlation function.  The highly
non-Gaussian nature of real images does not play a role in the estimation
problem at low SNR.

\subsection{Saddle point evaluation}
\label{ss:saddle}

In order to proceed, we need to be more specific about the distribution
of trajectories. A natural choice is that $\theta(t)$ undergoes Brownian
motion characterized by a diffusion constant $D$,
\begin{equation}
P[\theta(t)]=\frac{1}{Z}\exp\left[-\frac{1}{4D}\int\!\!
dt'\thd^2(t')\right].
\end{equation}
If we assume that the temporal pre-filter $T(\omega)$ is invertible, we can
use the variables $U_n(t)$ defined by
\begin{equation}
U_{n}(t)=\int\!\! T^{-1}(\tau)V_{n}(t-\tau).
\end{equation}
Finally, we will neglect the effect of aliasing and replace the discrete
photoreceptor lattice by a continuum, $U_n(t) \rightarrow U(x,t)$.  It is
useful to write things using the variables $t$ and $k$ conjugate to $x$.
In those variables the coupling term $\bar{U}(k,t)$ is related to $C(k,t)$
by a time dependent phase,
\begin{equation}
\bar{U}(k,t)=M(k)C(k,t)e^{ik\theta(t)}.
\label{eq:ubar}
\end{equation}

If the voltage noise is negligible, then the exponential in (\ref{eq:thde})
becomes a delta function and the functional integration over $C(k,t)$ amounts
to replacing $C(k,t)$ by $U(k,t)e^{-ik\theta(t)}/M(k)$, in other words
we simply forget about the voltage noise and write $U(k,t)=\bar{U}(k,t)$. To
write the form of the estimator we still need a more specific distribution
of contrast. We will assume that the action for $C(x,t)$ is local
in time and contains
no more than a quadratic term in time derivatives,
\begin{equation}
P[C(x,t)]=\frac{1}{Z}\exp\left[-\int\!\!dx\,dt\:\frac{\alpha}{2}\,
\dot{C}^2(x,t)
+\Gamma(C,\partial_xC,\partial_x^2C,\ldots)\right].
\end{equation}

With all these assumptions, we can write down the conditional probability
distribution for $\theta(t)$,
\begin{eqnarray}
P[\theta(t)|U(x,t)]&=&\frac{1}{Z}\exp\left[-\frac{1}{4D}\int\!\! dt \thd^2
\right.\nonumber \\
&&\qquad\mbox{}-\left.\frac{\alpha}{2}
\int\!\! dt \frac{dk}{2\pi} \frac{2ik \dot{U}^*U\thd
+k^2|U|^2\thd^2}{|M|^2}\right].
\end{eqnarray}
The potential term $\Gamma(C,\ldots)$ drops out because it is invariant
under a time dependent coordinate change ($x\rightarrow x+\theta(t)$) that
gets rid of the dependence on $\theta(t)$.
And the average value of $\thd(t)$ is given by the saddle point equation
\begin{equation}
\thd_{e}(t)=
{
{
\int dx \partial_x W(x,t) \partial_t W(x,t)
}
\over
{
D^{-1}+\int dx [\partial_x W(x,t)]^{2}
}
},
\label{eq:est-noiseless}
\end{equation}
where
\begin{equation}
W(x,t)=\int {{dk}\over{2\pi}} e^{ikx}\sqrt{{{2\alpha(k)}\over{| M (k)|^{2}}}}
\int dx' e^{-ikx'}U(x',t) .
\label{eq:est-noiseless2}
\end{equation}
In the last equation we have allowed $\alpha$ to depend on $k$, as it is the
case in the specific example we will consider in the following section.

One can easily understand equation (\ref{eq:est-noiseless}) intuitively.
For a fixed pattern $F(x)$ moving at a velocity $\dot{x}_\circ(t)$, the
pattern seen by an observer at rest would be $V(x,t)=F(x-x_\circ(t))$.  One
could compute back the velocity by the formula
$\dot{x}_\circ(t)={\partial_tV}/{\partial_xV}$.  So to estimate a velocity,
one can correlate spatial and temporal derivatives and normalize by spatial
derivatives.  If the pattern is itself changing in time, as a non-zero
$\alpha(k)$ would produce, then there would be spurious motion coming from
the variations in the image itself.  The optimal strategy in the presence
of such noise is, as always, to reduce your estimate by something like SNR
(remember (\ref{eq:toy-est})).  In this case the average ratio of spurious
motion to real motion is related to the product of the two constants $D$
and $\alpha$.

If the voltage noise is non-zero, equation (\ref{eq:est-noiseless}) will
have to be modified.  Since usually noise has a much shorter correlation
time than the images, it improves the performance to integrate the signal
over time to beat down the noise.  The issue is a little bit more subtle
than this since in our model the velocity also has zero correlation time
but it induces a correlation time for each spatial frequency of the images
equal to $1/k^2D$.  One might expect that the optimal estimator at any SNR
would look like a smeared out version of (\ref{eq:est-noiseless}).  We will
find something that hints in that direction when we consider a slightly
more specific model in the next section.  To find corrections to
(\ref{eq:est-noiseless}), one should not approximate the exponential in
(\ref{eq:thde}) and do a semi-classical approximation directly (which
amounts to finding the most probable $\theta(t)$, in itself a valid
estimator).  Unfortunately when one wants to go beyond the $N(\omega)=0$
limit, one is faced with integro-differential equations which do not lead
to any simple closed form expression for $\thd_e(t)$ as a functional of
$U(x,t)$.  We spare the reader more details on these fruitless efforts.

\subsection{Specific model: Gaussian contrast and white noise}
\label{ss:specific}

In this section, we will give a particular form to the noise power spectrum
and to the contrast probability distribution. This will allow us to
compute each term in the low SNR expansion and to have a better understanding
of what happens at higher SNR. For instance, in this model we can
show that, in the noiseless limit, the perturbation expansion matches up
term by term with the semi-classical result.

The effective contrast noise power spectrum will be taken to be white, i.e.
 $N(\omega)/|T(\omega)|^2$ equals a constant ($R^{-1}$).  If photon
shot-noise is the dominant source of noise, then this is the case and $R$
is the effective photon counting rate.  The white noise assumption allows
us to write the action locally in time and therefore to get rid of the
$\theta$-dependent phase (\ref{eq:ubar}) of the first term in the
exponential in Eq.  (\ref{eq:thde}),
\begin{eqnarray}
\thd_e(t)&=&\frac{1}{Z}\left\langle\thd(t)\exp\left[-\frac{R}{\phz}
\int\!\! dt'\int\!\!\frac{dk}{2\pi}
|M(k)C(k,t')|^2
\right.\right.\nonumber\\
&&\quad\left.\left.
\mbox{}+2U(k,t')M(k)C^*(k,t')e^{-ik\theta(t')}
\right]\right\rangle_{C,\theta}.
\label{eq:cgauss}
\end{eqnarray}

We need to specify a probability distribution for the contrast $C(x,t)$.
Picking a realistic one is by no mean an easy problem.  Even the statistics
of static images are hard to describe in simple terms \cite{scaling}.  We
saw in section \ref{ss:perturbation} that, to lowest order in SNR, only the
two-point function of $C(x,t)$ enters in the calculation.  Inspired by that
fact, we will use a Gaussian distribution for the probability of contrast.
Equation (\ref{eq:cgauss}) adds two more terms to the effective action for
$C$.  The first is uncoupled and quadratic; therefore it only modifies the
propagator without introducing any non-Gaussian interactions.  The second
couples $C$ linearly to $MUe^{-ik\theta}$.  Choosing $P[C]$ to be Gaussian
will allow us to do the $C$ integral exactly.  We will take the
spatio-temporal power spectrum of contrast to be of the form
\begin{equation}
S(k,\omega )
= S(k) {{2\tau(k)}\over {1 + [\omega\tau (k)]^2}} ,
\label{eq:contrastpower}
\end{equation}
which corresponds to a world where contrast fluctuations on an angular
scale $1/k$ have a variance $\sim S(k)$ and remain correlated for a time
$\tau(k)$.

We can now do the functional integral over $C(x,t)$ to find
\begin{equation}
P[\theta(t)|U(x,t)]=\frac{1}{Z}\exp\left[-\frac{1}{4D}\int\!\!dt\;\thd^2(t)-
H[\theta(t);U(x,t)]\right],
\label{eq:p-theta-eff}
\end{equation}
with the effective Hamiltonian given by
\begin{eqnarray}
H[\theta(t);U(x,t)]&=&-\frac{R^2}{2\phz^2}\int\!\! dt_1\int\!\! dt_2
\int\!\!\frac{dk}{2\pi}\left\{\sct(k)U(k,t_1)U(-k,t_2)\right. \nonumber \\
&&\qquad \times\left. e^{-|t_1-t_2|/\tilde{\tau}(k)}
e^{-ik[\theta(t_1)-\theta(t_2)]}\right\}.
\label{eq:hamiltonian}
\end{eqnarray}
where
\begin{equation}
\sct(k)=|M(k)|^2S_{C}(k)
\left\{1+\frac{2R\left|M(k)\right|^{2}S_{C}(k)\tau(k)}{\phz}\right\}^
{-\frac{1}{2}}
\label{eq:sctilde}
\end{equation}
\begin{equation}
\tilde{\tau}(k)= \tau(k)
\left\{1+\frac{2R\left|M(k)\right|^{2}S_{C}(k)\tau(k)}{\phz}\right\}^
{-\frac{1}{2}}.
\label{eq:tautilde}
\end{equation}
The second term within the braces in Eq.  (\ref{eq:sctilde}) and
(\ref{eq:tautilde}) is the $k$-dependent photoreceptor signal-to-noise
ratio.  $\tilde{\tau}(k)$ is the relevant time constant at spatial
frequency $k$.  It measures how for back in the past and in the future
extends the effect of the photoreceptor data on the optimal trajectory, or
in simpler terms, how far in time should one look to be able to compute the
optimal trajectory.  This constant will appear at every term in
perturbation theory and would appear in the exact result, if only it could
be written in closed form\ldots What is important, though, is that
$\tilde{\tau}(k)$ decreases with SNR.  It starts as $\tau(k)$ when the SNR
is very low.  At very low SNR, it is reasonable to integrate for as long as
the images are correlated to average over the noise, even at the expense of
a loss in high frequency resolution.  $\tilde{\tau}(k)$ goes to zero when
the SNR becomes very large.  As we saw in Section \ref{ss:saddle}, the
noiseless estimator is local in time.  Estimating a velocity in the absence
of noise amounts to correlating temporal and spatial derivatives, and
derivatives are local in time.  At intermediate SNR the optimum lies
somewhere between these two extremes.

It is worth noting that the SNR dependence of $\tilde \tau$ has a simple
interpretation.  In the case of linear filtering to recover a signal from a
noisy background (Section 2.2), we saw that the optimal reconstruction
filters the detector outputs so as to give near unit gain to frequencies
where the signal-to-noise ratio is high and then rolls off to suppress
frequencies where the signal to noise ratio is low [Eq.  (4)].  Roughly
speaking, the optimal reconstruction filter has a cutoff frequency at the
point where the signal-to-noise ratio crosses unity.  In our model where
the power spectrum of the signal is $S = |M(k)|^2 S(k ,\omega)$ [from Eq.
(46)] and the noise spectrum is $N= \phi_0 /R$, it is easy to see that for
large SNR the cutoff frequency when $S/N = 1$ is just $\omega_c =
\tilde\tau^{-1}$.  Thus a filter with time constant $\tilde\tau$ provides a
cutoff which rejects frequencies with less than unit signal-to-noise ratio.
 As the overall SNR is increased, one can trust the photoreceptor signals
out to higher frequencies.

Since we have an explicit expression for the conditional probability
distribution, one might hope that we could find its maximum which would
give us the maximum likelihood estimator.  What is involved is solving
\begin{equation}
\ddot{\theta}(t)=-2D\frac{\delta H[\theta(t)]}{\delta\theta(t)}.
\label{eq:integrodif}
\end{equation}
This integro-differential equation could in principle be solved, but in
practice there is no nice closed-form solution for general $U(x,t)$.  The
limit $R\rightarrow\infty$ of (\ref{eq:integrodif}) reduces to
(\ref{eq:est-noiseless}) with $\alpha(k)=2\tau(k)/S_c(k)$.

As we did in Section \ref{ss:perturbation}, we can calculate the optimal
estimator as a perturbation series is SNR by expanding the interaction
term, the exponential of the Hamiltonian (\ref{eq:hamiltonian}).  Here the
perturbation is a monomial, bilinear in $U$, so the $n^{\mbox{\scriptsize
th}}$ term in the series will have exactly $2n$ powers of $U$.  If we think
of the exact result as a functional of $U(x,t)$, then this series is also
the Volterra series in $U$ of that functional.  For example, the first
term, quadratic in $U$, can be viewed either as an approximation to the
optimal estimator at low SNR or as the quadratic part (in the Volterra
sense) of the exact result.

To compute the perturbation series, we expand in a Taylor series the
interaction $e^{-H}$ and take expectation values within the unperturbed
theory,
\begin{equation}
\thd_e(t)=
-\left\langle\thd(t)H(t)\right\rangle+
\frac{1}{2}\left\langle\thd(t)H(t)H(t)\right\rangle
-\left\langle\thd(t)H(t)\right\rangle
\left\langle H(t)\right\rangle+\ldots
\end{equation}
The last term comes from expanding the normalization $\left\langle e^{-H}
\right\rangle$. To compute the first term, we use the following identities:
\begin{equation}
\left\langle e^{-ik[\theta(t_1)-\theta(t_2)]}\right\rangle=
e^{-\frac{k^2}{2}\left\langle(\theta(t_1)-\theta(t_2))^2\right\rangle}
=e^{-k^2D|t_1-t_2|},
\end{equation}
\begin{eqnarray}
\left\langle\thd(t)e^{-ik[\theta(t_1)-\theta(t_2)]}\right\rangle&=&
\int\!\!dt'\left\langle\thd(t)\thd(t')\right\rangle
\left\langle\frac{\delta}{\delta\thd(t')}
e^{-ik[\theta(t_1)-\theta(t_2)]}\right\rangle\nonumber\\
&=& -2iDk[H(t_1-t)-H(t_2-t)]e^{-k^2D|t_1-t_2|}.\label{eq:gaussidentity}
\end{eqnarray}
With those tools in hand, we find the quadratic piece of the optimal estimator.
\begin{equation}
\thd_{e}^{[2]}(t)=\frac{2DR^{2}}{\phz^{2}}\int\frac{dk}{2\pi}ik
\sct(k)\int_{-\infty}^{t}\!\!\!\!dt_{1}\!\int_{t}^{\infty}\!\!\!\!dt_{2}U(k,t_1)
U(-k,t_{2})e^{-a(k)(t_2-t_1)}
\label{eq:thquad}
\end{equation}
where $a(k)=k^{2}D+\tilde{\tau}^{-1}(k)$. Equation (\ref{eq:thquad})
describes an antisymmetric correlator. In Section \ref{s:comparison}, we
will discuss in more details the meaning of this result.

Higher order terms in the perturbation series are easily computable using
identities like (\ref{eq:gaussidentity}). It is not difficult to write an
expression for the general term of this series, it is only cumbersome and
not very enlightening.
The quartic term contains all of the qualitative features of the following
terms (except of course that they will have more factors of $U(x,t)$).
It is given by
\begin{eqnarray}
&&\thd_{e}^{[4]}(t)=\frac{DR^{4}}{\phz^{4}}\int\frac{dk_{1}}{2\pi}(ik_{1})
\sct(k_{1})\int\frac{dk_{3}}{2\pi}
\sct(k_{3})\nonumber\\
&&\quad\mbox{}\times\int_{-\infty}^{t}\!\!dt_{1}\int_{t}^{\infty}\!\!dt_{2}
\int_{-\infty}^{\infty}\!\!dt_{3}
\int_{-\infty}^{\infty}\!\!dt_{4}U(k_{1},t_{1})
U(-k_{1},t_{2})U(k_{3},t_{3})U(-k_{3},t_{4})\nonumber\\
&&\quad\mbox{}\times\exp\left[-a(k_1)t_{21}
-a(k_3)t_{43}\right\}\left(\exp\left\{k_{1}k_{3}Dt_{1234}
\right]-1\right)\label{eq:thquart}
\end{eqnarray}
where $t_{ij}=|t_i-t_j|$ and $t_{1234}=t_{31}-t_{32}-t_{41}+t_{42}$.
Note that $|t_{1234}|$ is the overlap between the intervals $[t_1,t_2]$ and
$[t_3,t_4]$ and is zero if the two intervals don't overlap. The values of
$\{t_i\}$ that contribute are such that $t_1$ and $t_2$ are not too far from
$t$ (because of the exponential weighting), $t_3$ not too far from $t_4$ and
$[t_1,t_2]$ overlapping with $[t_3,t_4]$. So all the $\{t_i\}$ must be close
to $t$. The condition for overlapping intervals comes from the subtraction of
the disconnected diagrams and it happens at all orders in the expansion. That
means that if $\tilde{\tau}(k)$ goes to zero, as it does when SNR becomes
large, these terms should become localized in time.
If we let $R\rightarrow\infty$ in (\ref{eq:thquad}) and (\ref{eq:thquart})
we find
\begin{eqnarray}
\lim_{R\rightarrow\infty} \thd_e(t)&=&
D\int\frac{dk}{2\pi}ik\frac{U\dot{U}^*\tau}{|M|^2S_c} \nonumber \\
&&\mbox{}-D^2\int\frac{dk}{2\pi}ik\frac{U\dot{U}^*\tau}{|M|^2S_c}
\int\frac{dk'}{2\pi}(k'^2)\frac{|U|^2\tau}{|M|^2S_c}
+O[U^6].
\end{eqnarray}
This is the Volterra series of the noiseless estimator that we computed
earlier (\ref{eq:est-noiseless}) with $\alpha(k)=\tau(k)/2S_c(k)$.  This
strongly suggests that our model can be solved by perturbation theory.

With the help of (\ref{eq:est-noiseless}) we get a better understanding of
the higher order terms in the expansion.  The quartic term can be
interpreted as the beginning of saturation, i.e.  if the contrast is high
enough the estimator's response is no longer proportional to the square of
the contrast.  The response to a step displacement (delta function of
velocity) of the quartic term is opposite to that of the quadratic term.

\subsection{Causal velocity estimation}
\label{ss:causalvelo}

The low SNR estimator that we found in the previous section
(\ref{eq:thquad}) involves both the photoreceptor data from the past and
from the future.  A real fly, or any other physical realization of the
estimator, might not be able to afford such a luxury.  We might be tempted
to simply truncate the $t_2$ integral to only allow for a small delay $T$
in the estimation.  If $T$ is large compared with the time constants
$1/a(k)$ then this procedure should be very close to optimal.  But in
general, this might not be optimal.

To fit in the constraint of causality in our computation, we would like to
do as in Section \ref{ss:causal}: solve the acausal problem and average
over the conditional distribution of the unknown data given the observed
data. The Wiener factorization trick only works for Gaussian distributions,
or equivalently for linear filters, and is not easily generalizable to
arbitrary distribution. Fortunately the quadratic estimator
(\ref{eq:thquad}) is actually linear in the future data. Averaging over the
values of $U(x,t')$ for $t'>t+T$ will then amount to replacing $U(x,t')$ by
its average value in the conditional probability $P[U^+|U^-]$. This can be
obtained by Wiener filtering and requires only knowledge of the power
spectrum for $U(x,t)$.
\begin{equation}
\left\langle U(k,t)U(k',t')\right\rangle=
2\pi\delta(k+k')\left\{\frac{\phz}{R}\delta(t-t')+S_c(k)e^{-(k^2D+\tau^{-1}(k))
|t-t'|}\right\}
\end{equation}
As described in Section \ref{ss:causal} we write the reciprocal of this
power spectrum as $|\Psi_k(\omega)|^2$ where $\Psi_k(\omega)$ doesn't have
any singularities in the upper half plane.  The index $k$ just goes along
for the ride.  We then write the ``equation of motion'' for the average
value of $U^+$ given $U^-$ (\ref{eq:eqofmotion}).  It is an integral
equation whose solution is given by
\begin{equation}
U(k,t)=(b-\beta)\left[\int_{-\infty}^0 dt'U(k,t')e^{\beta t'}\right]e^{-bt}
\mbox{ for } t>0,
\label{eq:u+}
\end{equation}
where
\begin{equation}
b(k)=k^2D+\tau^{-1}(k)
\end{equation}
and
\begin{equation}
\beta(k)=b(k)\sqrt{1+\frac{2R|M(k)|^2S_c(k)}{b(k)\phz}}.
\end{equation}
We can now substitute (\ref{eq:u+}) into (\ref{eq:thquad}) evaluated at
$t=-T$.  Using time translation invariance we can then write the causal
estimator evaluated at any time.
\begin{eqnarray}
&&\thd_{eC}^{[2]}(t)=\frac{2DR^{2}}{\phz^{2}}\int\frac{dk}{2\pi}ik
\sct(k)\left\{\int_{-\infty}^{t}\!\!\!\!dt_{1}\!\int_{t}^{t+T}\!\!\!\!dt_{2}
U(k,t_1)U(-k,t_{2})e^{-a(k)(t_2-t_1)}\right. \nonumber\\
&&+\left.\frac{b(k)-\beta(k)}{a(k)+b(k)}
\int_{-\infty}^{t}\!\!\!\!dt_{1}\!\int_{-\infty}^{t+T}\!\!\!\!dt_{2}
U(k,t_1)U(-k,t_{2})e^{a(k)(t_1-t-T)-\beta(k)(t_2-t-T)}\right\}
\label{eq:est-causal}
\end{eqnarray}
To lowest order SNR, the different time constants $a(k)$,$b(k)$ and
$\beta(k)$ are all equal and the second term vanishes leaving us with the
truncated estimator.  At higher SNR, it is not clear what this term should
become. If $T$ were zero, it would be impossible to estimate the velocity
since it has delta function correlations.  When we put $T=0$ in the second
term, we get identically zero from the anti-symmetry in $k$ if we
identify $a(k)$ with $\beta(k)$; these are not strictly equal but their
difference might come from our uncontrolled approximations.  The main result
of this section is that the truncated estimator is the optimal causal
estimator at low SNR.

\section{Characterization}

Imagine that we are presented with a physical device which we suspect may
embody the optimal velocity estimator.  How can we probe the system to see
if our suspicions are correct?  What are the key experimental signatures of
the optimal estimator?  One idea is that the optimal estimator is different in
different regimes of SNR, so if the putative optimal estimator
is designed to adapt to a wide range of conditions we might hope to trap the
system in these different regimes and observe different responses to the same
input signals.  These `adapt and probe' experiments \cite{rob-H1adapt} are the
analog of the standard pump-probe experiments in condensed matter physics.
Another point is that the optimal estimator is, by definition, as reliable as
possible given the input signals.  We can try to quantify this reliability by
directly measuring the noise in the estimator and then referring this noise
back to the input as an effective input velocity noise.  This is a standard
procedure for characterizing noise in electronic devices
\cite{horowitz+hill} and is also used to quantify the
performance of neurons in real-time estimation tasks \cite{science}.

\subsection{Response function}

Let the optimal estimator be set up to match a world of given statistical
characteristics (power spectrum, photon flux, ...).  Rather than choosing
contrasts $C(x,t)$ from this distribution, we want to probe the system with
some stereotyped stimulus.  For simplicity we choose a static random
pattern with spatial spectral density $S_C (k)$ and we have this pattern
make a small step displacement at time $t=0$.  We are interested in the
experimentally observable average response of the causal estimator
(\ref{eq:est-causal})).  By average, we mean here an average over the noise
and over realizations of the pattern.  The angular trajectory for a step at
$t=0$ is simply $\theta(t)=\bar{\theta} H(t)$, $H(t)$ being the Heaviside
step function and $\bar{\theta}$ a small angle compared to $\phz$ and to
the width of the photoreceptors optical profile.

Since the noise has delta function correlation and the estimator doesn't
correlate voltages at the same time, there is no noise contribution to the
average response.  All we need to do is the contrast average and the
$t$-integrals.
\begin{eqnarray}
\thd_e^{\mbox{\scriptsize res}}(t) & = &
\frac{2DR^2}{\phz^2}\int\!\!\frac{dk}{2\pi}
ik\sct(k)
\int_{-\infty}^{t}\!\!dt_1
\int_{t}^{t+T}\!\!dt_2
|M(k)|^2
\left\langle C(k,t_1)C(k,t_2)\right\rangle \nonumber \\
& & \qquad\times \exp\left[ik\bar{\theta}(H(t_1)-H(t_2))\right]
e^{-a(k)(t_2-t_1)}
\end{eqnarray}
We are interested in the small $\bar{\theta}$ response, so we
will expand the exponential of $ik\bar{\theta}$ up to the first
non-zero term. Substituting the average over $C^2$ by its power
spectrum and performing the $t$-integrals we find
\begin{equation}
\thd_e^{\mbox{\scriptsize res}}(t)=\frac{2NDR^2\bar{\theta}}{\phz}
\int\!\!\frac{dk}{2\pi}\frac{k^2|M(k)|^2\sct(k)S_C(k)}{a^2(k)}
\left\{
\begin{array}{l}
0\\
e^{a(k)t}-e^{a(k)T}\\
\left(1-e^{-a(k)T}\right)e^{-a(k)t}.
\end{array}
\right.\label{eq:resp}
\end{equation}
The three cases being respectively $t<-T$, $-T<t<0$ and $t>0$.
Unfortunately the last $k$-integral can't be done analytically, but still
the last equation gives us a good idea of the response.  If $T$ is short
then the response rises very rapidly just before the step.  Note that in a
real implementation the rise would occur between the step and the delay
time $T$. After the step, the response decays with a time constant related to
$\taut(k)$. The response to the acausal quadratic estimator
(\ref{eq:thquad}) is also given by (\ref{eq:resp}) with
$T\rightarrow\infty$. In other words, it is symmetric about zero and each
$k$ component rises and decays with time constant $1/a(k)$.

At high SNR, the integrated response is independent of the
contrast power spectrum $S_C(k)$.  One can see this by looking at the time
integral of (\ref{eq:resp}) and noticing that it contains three powers of
$\taut(k)$ and one of $\sct(k)$ whose renormalization factors
(\ref{eq:sctilde}) cancel the two powers of $S_C(k)$.

\subsection{Equivalent input noise}

To quantify the performance of an estimator, we need to define a notion of
noise in the estimate.  We would like to say that the estimate is
equivalent to the true signal to which noise has been added.  This added
noise corresponds to the combination of the external noise and of the noise
introduce by the estimation process.  Since least-mean-square estimators
underestimate the signal, we want to correct for systematics.
For example in the toy-model considered in Section \ref{s:simple}, Eq.
(\ref{eq:toy-est}), the external noise has variance $N$ but the chi-square
of the estimate is $N/(1+N/S)$.  We have to first find the gain between the
estimate and the signal and then compute the power spectrum of the
difference between the signal and the estimate divided by the gain.  In the
engineering literature this quantity is referred to as the {\em equivalent
input noise power}.  For time dependent signals, the gain is a linear
filter and is easily dealt with in frequency space.  In the case at hand
velocity is the signal. We write
\begin{equation}
\thd_e(\omega)=g(\omega)\left(\thd(\omega)+\eta(\omega)\right),
\end{equation}
\begin{equation}
\mbox{ where $g(\omega)$ minimizes }
\left\langle\left|\thd_e(\omega)-g(\omega)\thd(\omega)\right|^2\right\rangle
{}.
\label{eq:noisedef}
\end{equation}
We are interested in the power spectrum of $\eta$ which we find using
the definition (\ref{eq:noisedef}) and some algebra,
\begin{equation}
\left\langle|\eta(\omega)|^2\right\rangle=
\left\langle|\thd(\omega)|^2\right\rangle
\left\{
\frac{
\left\langle|\thd(\omega)|^2\right\rangle
\left\langle|\thd_e(\omega)|^2\right\rangle
}{
\left|\left\langle\thd(\omega)\thd_e^*(\omega)\right\rangle\right|^2
}
-1\right\}.
\label{eq:noisepower}
\end{equation}
All these expectation values are proportional to
$2\pi\delta(\omega=0)$ which should be factored out when we
compute the power spectrum. For the toy-model (\ref{eq:toy-est}) the
variance of the equivalent input noise is equal to the external noise
variance $N$.

\subsubsection{High SNR estimator}
\label{ss:hsnrnoise}

The first estimator that we will characterize is the instantaneous
estimator (\ref{eq:est-noiseless},\ref{eq:est-noiseless2}) that was
obtained by neglecting voltage noise.  Although voltage noise is assumed to
vanish, the remaining signals are not perfect indicators of motion, since
the contrast $C(x,t)$ can fluctuate even if there is no motion; these
time-dependent contrasts set a limit to reliability of motion estimation
even in the absence of true detector noise.  As in Section
\ref{ss:specific} we will assume Gaussian contrast distribution with power
spectrum given by (\ref{eq:contrastpower}), with the only difference that
we will specify the function $\tau(k)$ to be a constant $\tau$.  The
estimator is then given by
\begin{equation}
\thd_e(t)=\frac{D\tau\int\!\!\frac{dk}{2\pi}ikW(k,t)\dot{W}(-k,t)}
{1+D\tau\int\!\!\frac{dk}{2\pi}k^2|W(k,t)|^2} \mbox{ where }
W(k,t)=\frac{U(k,t)}{\sqrt{S_C(k)|M(k)|^2}}. \label{eq:est-noiseless3}
\end{equation}

To compute the equivalent input noise using (\ref{eq:noisepower}) we need
to take expectation values of products of Gaussian fields some of which
appear in the denominator.  It is unfortunately impossible to do so exactly
but we can expand those operators in their fluctuations around their
average value, the former being typically smaller than the latter by a
factor of $N$, the number of photoreceptors.  The factors of $N$ will enter
the calculation when we take the expectation value of the contrast
evaluated at $k$ and $-k$.  Formally this would be proportional to
$2\pi\delta(k=0)$ because we have used an infinite size approximation.  For
finite sample size, it should be proportional to the size of the system
($N\phi_\circ$).  Another problem is the divergent $k$-integrals for which
we impose a hard cut-off at the sampling frequency ($k=\pm\pi/\phi_\circ$).
After some straight-forward computations we find the input noise to first
order in contrast fluctuations
\begin{equation}
N_{\mbox{\scriptsize input}}(\omega)=
\frac{18D}{5N}\left[1+
\left(\frac{2\pi^2ND\tau}{3\phi_\circ^2}\right)^2
\right]^{-1}+
\frac{3\phz^2\tau\omega^2}{\pi^2N(4+\tau^2\omega^2)}.
\label{eq:noisehsnr}
\end{equation}
The two terms in the power spectrum of the noise are, as stated above, the
respective contribution of the overall fluctuations in mean-square contrast
and of the time-dependent changes in the image.  If the correlation time of
the images is not too short ($\phi_\circ^2/D\tau$ small compare to $N$)
then the first term goes like $1/N^3$ and should be neglected to be
consistent with our approximations.  On the other hand if we only consider
the numerator of (\ref{eq:est-noiseless3}), corresponding to a correlator
without saturation, we can redo the calculations leading to
(\ref{eq:noisehsnr}) and we find that the first term simply becomes
$18D/5N$.  This term would be present even in the absence of image
decorrelation ($\tau\rightarrow\infty$).  In other words, an unsaturating
correlator is always subjected to noise coming from overall contrast
fluctuations.

\subsubsection{Low SNR estimator}

The next estimator we want to characterize is the one given by the first
term in the low SNR expansion (\ref{eq:est-lowsnr}).  To be more specific
we will use equation (\ref{eq:thquad}) obtained by considering white noise
and Gaussian contrast.  To lowest order in SNR, the renormalized time
constant $\tilde{\tau}(k)$ and contrast spectrum $\sct(k)$ are equal to
their non-renormalized counterpart except for an obvious factor of
$|M(k)|^2$.  Since in this section we will limit ourselves to the dominant
noise contribution at low SNR, we will use the non-renormalized functions.
We use equation (\ref{eq:noisepower}) to compute the equivalent input noise
power spectrum, keeping only the dominant terms at low SNR.  In particular
we only include the noise-noise contribution of the estimator's
auto-correlation.  Schematically if
\begin{equation}
\thd_e=\int\!\! gVV
\end{equation}
then
\begin{equation}
\left\langle\thd_e\thd_e\right\rangle\approx
\left\langle\thd_e\thd_e\right\rangle_{NN}=
\int\!\! g_1g_2\left\langle\delta V_1\delta V_1\delta
V_2\delta V_2\right\rangle
\end{equation}
where $\delta V$ is the voltage noise. We also drop the $-1$ term
within the braces of (\ref{eq:noisepower}) since it is of higher
order in SNR. In this limit the calculation is rather simple
because the voltage noise correlations are delta functions of
time and the kernel $g$ is given by the correlation of $\thd$
and $VV$ so both $\langle|\thd_e|^2\rangle$ and
$\langle\thd\thd_e^*\rangle$ are proportional to the integral
of $g^2$. If we compute these correlations precisely we get
\begin{equation}
\langle\thd_e(t)\thd_e(0)\rangle=
\frac{ND\phi_\circ}{2}\int\!\!\frac{dk}{2\pi}
\frac{k^2|M(k)|^4S_C^2(k)}
{b^2(k)}
e^{-2b(k)|t|},
\end{equation}
where $b(k)=k^2D+\tau^{-1}(k)$ and
with a similar expression for $\langle\thd\thd_e^*\rangle$.
So the equivalent input noise power is given by
\begin{equation}
N_{\mbox{\scriptsize input}}(\omega)=\frac{\phi_\circ}{NR^2}
\left\{\int\!\!\frac{dk}{2\pi}
\frac{k^2 |M(k)|^4 S_C^2(k)}{b(k)}
\frac{1}{4b^2(k)+\omega^2}\right\}^{-1}.
\end{equation}
Notice that the noise goes like $1/N$ and is proportional to the square
of the inverse effective photon counting rate. At low frequencies it is almost
constant and at high frequencies it goes like $\omega^2$. Thus although we
constructed the optimal estimator for the velocity $\thd$, at high
frequencies the effective noise level for estimating $\theta$ itself in
frequency-independent. This feature of the optimal estimator is observed
in experiments on flies \cite{science}.

\subsubsection{Renormalized correlator}

Finally we have shown in Section \ref{ss:specific} that the renormalized
correlator (\ref{eq:thquad}) is the quadratic part of the Volterra series
of the exact optimal estimator.  One might want to take this estimator more
seriously than just a low SNR estimator and see what its performance is at
any SNR.  To do so, one just has to compute all the correlation functions
in the noise formula (\ref{eq:noisepower}).  This is a rather
straight-forward computation.  We actually did it at any frequency
($\omega$), but the results were not easily presentable so we decided to
only include them for $\omega=0$, and even then, we find it more proper to
put them in the Appendix.  There are a few points that are worth making.
First, the low SNR result of the previous section can be recovered by
considering only $A_4$ and $A_0$ in (\ref{eq:noiseappendix}) and letting
$\sct(k)\rightarrow |M(k)|^2S_C(k)$ and $\taut(k)\rightarrow \tau(k)$.
Secondly, there is a term, namely $A_1$, which doesn't decrease with the
sample size $N$.  As the SNR gets large and the correlator becomes
localized in time, this term shrinks.  It disappears in the SNR goes to
infinity limit and we recover the result mentioned in Section
\ref{ss:hsnrnoise} for a noiseless optimal correlator.

The fact that Eq.  (\ref{eq:noiseappendix}) has a term which is independent
of $N$ is a bit strange.  Intuitively we are looking for a motion signal
which is coherent across all $N$ receptors, while noise sources are local
and incoherent across the array, so that the optimal estimator must have a
noise level which decreases as $1/N$.  In fact one can prove this quite
trivially by assuming that the estimated trajectory $\theta (t)$ is close
to the true trajectory and expanding the effective Hamiltonian [Eq.  (48)];
the ``stiffness'' of the system is extensive in $N$ and hence the effective
noise level is $\propto 1/N$.  What then is wrong with the renormalized
correlator? We saw that the correlator is just the first term in an
expansion of the optimal estimator, and that this term is guaranteed to
dominate only at low SNR.  Indeed, at low SNR the correlator gives an
effective noise level with the appropriate $N$ dependence [Eq.  (73)], but
evidently the higher order terms play a crucial role in maintaining the
correct $N$ dependence at intermediate SNR.  We now show that this is a
symptom of a more general result: any correlator that integrates
photoreceptor voltages for some finite time will generate a systematic
error in its velocity estimate which cannot be undone by linear filtering,
and hence an $N-$independent term in the effective noise level.

A general correlator has the form
\begin{equation}
\thd_e(t)=\int g(k,t_1-t,t_2-t)V(k,t_1)V(-k,t_2).
\label{eq:gencor}
\end{equation}
The term in the noise (\ref{eq:noisepower}) with the highest power of $N$
arises when one considers  Wick contractions of the contrast field within the
same $\thd_e$, which amounts to computing the noise in the
contrast-averaged estimator. Note that any reasonable correlator
has no average voltage noise contributions.
\begin{equation}
E(\theta;t)=\left\langle\thd_e(t)\right\rangle_C=
N\phz\int g(k,t_1-t,t_2-t)S_C(k,t_{21})e^{ik[\theta(t_1)-\theta(t_2)]}
\end{equation}
The point is that the function $E$ is not linearly related to the
velocity $\thd$, which means that there is a systematic discrepancy
between the estimator and the true velocity. To finish the proof, we make
use of the following identity for functions of Gaussian variables:
\begin{equation}
\langle F(X)^2\rangle=\langle F(X)\rangle^2+\int\!\! dtdt'\langle
X(t)X(t')\rangle\langle\frac{\delta F}{\delta X(t)}\frac{\delta F}{\delta
X(t')}\rangle,
\end{equation}
where $X(t)$ is a Gaussian variable with zero mean. From this we can
derive the following identity for the function $E(\theta;t)$:
\begin{equation}
\int\!\!dt\left\langle E(0)E(t)\right\rangle=
2D\int\!\!dtdt'\left\langle\frac{\delta E(t)}{\delta \thd(0)}
\frac{\delta E(t')}{\delta \thd(0)}\right\rangle.
\end{equation}

In the last expression, we have used the explicit two-point function of
$\thd$ and the fact that $E$ averaged over all trajectories should be
zero by symmetry. We have also used time translation invariance to shift
the $t$-integrals.
The $N$-independent term in the noise at zero frequency is then
\begin{equation}
N_{\mbox{input}}(\omega=0)=2D\frac{\left\langle(\Delta G)^2\right\rangle}
{\left\langle G \right\rangle^2}
\mbox{ where }
G=\int\!\! dt \frac{\delta E(t)}{\delta \thd(0)}.
\end{equation}
We find the general result that the $N$-independent noise piece is given
by the normalized variance of the function $G$, which is the non-linear
gain of the estimator as a function of velocity. In the case of a general
correlator (\ref{eq:gencor}), $G$ is given by
\begin{equation}
G=N\phz\int\!\!\frac{dk}{2\pi}dt_1dt_2
ikg(k,t_1,t_2)S_C(k,t_{21})
\int_{0}^{t_{21}}\!\!\!dte^{ik[\theta(-t)-\theta(t_{21}-t)]}.
\end{equation}
In general $G$ will depend functionally on $\theta(t)$ and therefore will
have a non-zero variance. The only case where $G$ can be independent of
$\theta$ is if $g(k,t_1,t_2)$ is zero for $t_1\neq t_2$, i.e. the
correlator is instantaneous. This complete our proof that any correlator
with non-zero support has an $N$-independent noise piece. Although we
don't have a formal proof of this, it appears that this systematic error
cannot be undone by a time independent non-linearity at the output.

\section{Comparison with experiment}
\label{s:comparison}
Let us suppose that the fly's visual system in fact embodies the optimal
estimator whose structure has been discussed in the previous sections.  How
would we compare theory and experiment? We recall that the fly's estimate
of angular velocity can be probed both in the behavior of the whole fly and
in the response of individual motion-sensitive neurons.  In each case one
measures something which has different units than the angular velocity
itself, so one must be careful in interpreting absolute quantities.  The
dependence of the response on various parameters of the stimulus, however,
should be a robust prediction of the theory.  Here we draw attention to
some of these robust features, and comment on the comparison to experiments
in the literature.

\begin{itemize}
\item At both high and low signal-to-noise ratios, an essential element of
motion estimation is the cross-correlation of photoreceptor outputs.  At
high SNR the correlation is between the spatial and temporal derivatives,
suitably normalized, while at low SNR results the correlation is between
spatially and temporally smeared versions of these derivatives.
\item The particular spatial and temporal filters which must be applied to
the photoreceptor signals before computing the cross-correlation depend on
the SNR, on the statistics of the trajectory $\theta (t)$ and on the
statistics of the visual world defined by $P[C(x,t)]$.  Thus the optimal
motion estimator is an {\it adaptive} correlator.
\item As the SNR becomes large, the optimal estimator crosses over smoothly
to being a comparison between spatial and temporal derivatives, as in Eq.
\ref{eq:est-noiseless} .  This smooth crossover is related to the fact that
our statistical mechanics problem does not have a phase transition, so that
in some sense the motion estimation problem is solvable in perturbation
theory.

\end{itemize}

Remarkably, all of these features of the optimal motion estimator have
correlates in experiments on real flies.

\subsection{Correlation}

The idea that insects estimate motion by a correlation scheme dates back
forty years, to the classic work of Hassenstein and Reichardt \cite{hass}.
There is an enormous body of evidence that fly optomotor behavior can be
described at least approximately by a correlation model
\cite{buchner84,quartrevII,quartrevI}, and the same can be said for the
responses of H1 and the other movement sensitive neurons
\cite{borst+egelhaaf-TINS,zaagman78}.  The key experimental test is the
demonstration that the response of the motion-sensitive system depends
quadratically on the stimuli to individual photoreceptors, and indeed this
is observed for low contrast stimuli.

Poggio and Reichardt \cite{quartrevII,quartrevI} tried to cast the problem
of motion estimation in a more general formalism, which consists
essentially of a Volterra series expansion of the functional which relates
the receptor signals to the estimator.  They emphasized that the 2nd order
correlation term is the simplest term in this series which can give an
estimate which is related to the real motion signal.  What was unclear in
this formulation is why the system should use the simplest term.  Further
we know that the system is not {\it exactly} described by the correlation
model, for example at high contrasts, so in this formulation it appears
that flies do something more complex than they ``need'' to do.  The
statistical mechanics approach shows us that the series expansion in the
Poggio-Reichardt work is justifiable as an expansion in signal-to-noise
ratio, that the first term dominates only at low SNR, but there is a
well-defined limit at high SNR which is also surprisingly similar to the
simple correlation scheme.  It is known that the impulse response of the
correlator model (\ref{eq:resp}) as the same qualitative features as the
response of the cell H1 to a step displacement
\cite{rob-H1adapt,srini-imp-res}.  Perhaps more importantly, the spatial
and temporal filters which are arbitrary from the earlier point of view are
completely determined if the fly is to build the optimal estimator.

\subsection{Adaptive filtering}

This brings us to our second point, namely that the spatial and temporal
filters must adapt to the statistics of the environment.  Correlations
should be computed not just between nearest neighbors but over some range
of distances which depends on the signal-to-noise ratio.  There is direct
evidence for distant neighbor correlation both in fly behavior
\cite{buchner76} and in the responses of H1 \cite{lenting-thesis,schuling},
and the relative weights of correlation at different distances changes with
background light intensity in qualitative accord with theory.  By measuring
the transient responses of H1 one finds that the time constants of the
filters which precede the correlation computation can adapt over a range of
more than one order of magnitude \cite{maddess,rob-H1adapt,zaagman83}, and
this adaptation is determined locally on the scale of at most a few
photoreceptor spacings, as would be required for optimal processing in an
inhomogeneous environment.  To date this adaptation has been probed using
constant velocity motion of rigid patterns, and other deterministic
stimuli.  It is clear from the theory, however, that the optimal processor
should adapt its filtering to the correlation time and variance of random
input signals.  This is an important prediction because adaptation in the
nervous system is usually described as the gradual fading away of responses
to constant input, as in the familiar example of light adaptation where we
are first blinded by a bright light and gradually recover sensitivity to
small amounts of contrast around the large mean level; thus classical
neural adaptation involves adaptation to the first moment of the stimulus
distribution.  Here we predict that the system instead changes its dynamics
in response to a higher-order statistical feature, the correlation time.
The statistical framework we have proposed for signal processing makes the
clear qualitative prediction that the optimal processor must adapt to the
{\it ensemble} of input signals, not just to the mean level, and this
prediction is independent of almost all details.  It is thus important to
test for this {\em statistical adaptation} in real neurons.

\subsection{Beyond correlation}

Finally, the saturation behavior of the optimal estimator points to the
essence of current controversies in the literature on visual motion
detection.  As far as we know, most discussions of motion estimation begin
by pointing out that there are two qualitatively different approaches (see,
for example, Ref.  \cite{koch-hildreth}).  In the first, inspired by
experiments on insects, one correlates the responses of neighboring
receptors, while in the alternative scheme one compares spatial and
temporal gradients.  Certainly these seem like very different
algorithms---in one case the essential non-linear operation is
multiplication, in the other case division.  Furthermore, taking ratios of
gradients provides an invariance to changes in overall contrast or spatial
pattern, while Reichardt's original correlation scheme does not measure a
true velocity but rather confounds angular velocity with the contrast and
spatial structure of the image.  It is tempting to suggest that ``simpler''
animals (like flies) are limited in what they can compute, and therefore
use the correlation scheme despite its difficulties, while higher animals
(like us?) have more computational power and hence derive unambiguous
estimates of motion.  The experimental situation is far from clear.  Many
of the models designed to account for human perceptual performance can
actually be reduced to variations on the correlation model
\cite{koch-hildreth}, while at least one group
\cite{horridge-mult,horridge-contrast} has drawn attention to the aspects
of insect vision which do not conform to the correlation scheme and
suggested instead that flies do something more closely approximated by
gradient comparison model.  Finally, under statistically stationary
conditions one can decode the responses of the fly's motion sensitive
neuron H1 and recover an unambiguous if slightly noisy estimate of the
unknown, time varying angular velocity signal \cite{science}, indicating
that at least under these conditions the fly's brain does not confound
motion with other aspects of the visual world.  Can the theory of optimal
estimation help us resolve these apparent conflicts?

Perhaps the most important point is that correlation and gradient
comparison are {\it not} qualitatively different approaches to motion
estimation, but rather different limits of a smoothly varying family of
algorithms adapted to different signal-to-noise ratios.  If flies (or
humans, for that matter) perform optimal motion estimation in environments
which span a wide range of SNR, then they will sometimes appear to use the
correlation scheme and sometimes appear to use gradient comparison.  The
pure versions of these models are valid only at infinitesimal or at
infinite SNR respectively, and at least in the simple ``model worlds'' we
have considered the crossover between these limits is smooth.  Thus the
optimal motion estimator should show the classic quadratic dependence on
contrast in a low contrast (and hence low SNR) world, then smoothly
saturate to a contrast-independent response at high contrasts, provided
that the photon flux is large enough to allow the development of high SNRs.

The origin of the saturation behavior is actually two-fold.  If one looks
just at the 2nd order correlator term in the perturbation expansion, then
because the time constants and spectral densities are renormalized in
relation to the SNR, one finds that the average response to slow, constant
velocity motion in fact saturates as the spectral density of contrast
fluctuations in increased.  Examination of higher terms in the perturbation
expansion shows the same behavior, but as higher order terms become
important the series sums to give the gradient comparison.  The crucial
point is that saturation is not just a static cutoff of the quadratic
growth at low contrast, but rather a subtle combination of mechanisms in
which the coefficients of the optimal computation must be reset in relation
to the signal-to-noise ratio.  This means that the entire response {\it
vs.} contrast curve should change depending on the rms contrast to which
the system is adapted, that the deviation from quadratic response should be
associated with the onset of this adaptive behavior, and that this
crossover point where adaptation begins should correspond to an rms
contrast which provides an SNR near unity.

\subsection{Crossover and noise}

The details of the crossover from the correlator to gradient comparison may
seem uninteresting, but in fact these details are essential for insuring
that the reliability of the optimal estimator improves with the number of
photoreceptors.  We have seen that the a broad class of correlator models,
even adapting correlator models, fail to give an effective noise level
which declines with $N$, and hence are {\em qualitatively} sub-optimal.  We
know that under some conditions the noise level in the fly's velocity
estimate approaches the limit corresponding to the optimal estimator
\cite{santafe,science,fred-thesis,stepdisc-nips,rob-abs84}, including the
factor of $1/N$, but these tests are limited to relatively low SNR and the
$N$-dependence of the effective noise level has not been probed directly.
Since the higher-order terms in the series (\ref{eq:perturbation}) play the
decisive role in setting the $N$-dependence of the noise level one might
hope to find a clear qualitative signature of these terms in the response
to properly chosen stimuli; this remains an open problem.

\subsection{Summary}

To summarize, the optimal motion estimator has many qualitative features in
common with the neural motion estimator found in the fly's brain.  Theory
suggests that even some apparently conflicting observations may be understood
in terms of the rich adaptive behavior of the optimal estimator.  This adaptive
behavior should be directly observable, and would constitute evidence for
adaptation to higher-order statistics rather than just adaptation to the DC
level.  Finally, there is the quantitative prediction that the onset of
adaptation to the signal-to-noise ratio and the concomitant departure from
quadratic contrast dependence should begin at $SNR =1$.

\section*{Acknowledgments}
This work was inspired by the possibility of combining theory with direct
experimental tests in the context of fly vision, and we are grateful to R. de
Ruyter for offering us this challenge.  We also thank E. Hsu, A. Libchaber, M.
Meister, F. Rieke, D. Warland and M. Wexler for many helpful discussions.  Work
at Princeton University was supported in part by NSERC of Canada through a '67
award to M.P.

\appendix

\section*{Appendix}
For completeness we give here the exact equivalent input noise at zero
frequency in the
renormalized correlator (\ref{eq:thquad}).
\begin{equation}
N_{\mbox{\scriptsize input}}(\omega=0)=
A_0^{-2} \left [
A_1+\frac{1}{N}\left (A_2+\frac{A_3}{R}+\frac{A_4}{R^2}
\right ) \right ]
\label{eq:noiseappendix}
\end{equation}
\begin{equation}
A_0=\int\!\!\frac{dk}{2\pi}\frac{k^2|M(k)|^2S_C(k)\sct(k)}{[a(k)+b(k)]^3}
\end{equation}
\begin{eqnarray}
A_1&=&4D^3\int\!\!\int\!\!\frac{dk}{2\pi}\frac{dk'}{2\pi}
|M(k)|^2|M(k')|^2S_C(k)\sct(k)S_C(k')\sct(k')k^4k'^4\nonumber \\
&&\quad\times
\left\{
c^3(k)c^3(k')
\left[
(c(k)+c(k'))^2-4D^2k^2k'^2
\right]^3
\right\}^{-1}
\nonumber\\
&&\quad\times
\left\{
(c(k)+c(k'))^2
\left(
4c^2(k)+11c(k)c(k')+4c^2(k')
\right)
\right.\nonumber\\
&&\qquad \left.
\mbox{}-4D^2k^2k'^2
\left(
6c^2(k)+11c(k)c(k')+6c^2(k')+8D^2k^2k'^2
\right)
\right\}
\end{eqnarray}
\begin{eqnarray}
A_2&=&\frac{1}{4\phz}\int\!\!\frac{dk}{2\pi}
k^2|M(k)|^4S_C^2(k)\sct^2(k)\tau^{-1}(k) \nonumber\\
&&\quad\times\left\{
(a+b)^4(a+b+k^2D)^3(a+2k^2D)^3
\right\}^{-1}\nonumber \\
&&\quad\times\left\{
8a^4+21a^3b+19a^2b^2+7ab^3+b^4+33a^3k^2D+70a^2bk^2D
\right.\nonumber \\
&&\qquad\left.
\mbox{}+45ab^2k^2D+8b^3k^2D+46a^2(k^2D)^2+72ab(k^2D)^2
\right. \nonumber\\
&&\qquad\left.
\mbox{}+27b^2(k^2D)^2+24a(k^2D)^3+20b(k^2D)^3+4(k^2D)^4
\right\}
\end{eqnarray}
\begin{equation}
A_3=\frac{1}{4}\int\!\!\frac{dk}{2\pi}|M(k)|^2S_C(k)\sct^2(k)k^2
\frac{a^5+5a^4b+a^3b^2-4a^2b^3-4ab^4-b^5}{a^3(a+b)^4(a^2+b^2)}
\end{equation}
\begin{equation}
A_4=\frac{\phz}{16}\int\!\!\frac{dk}{2\pi}\frac{k^2\sct^2(k)}{a^3(k)}
\end{equation}
where $a(k)=k^2D+\taut^{-1}(k)$, $b(k)=k^2D+\tau^{-1}(k)$ and $c(k)=a(k)+b(k)$.


\begin{thebibliography}{99}


%\bibitem{exp-tails:anselmet84}
%R. Antonia, E. Hopfinger, Y. Gagne, and F. Anselmet (1984).
%Temperature structure function in turbulent shear flows,
%{\it Phys. Rev. A} {\bf 30,}
%2704-2707.

%\bibitem{atickrev}
%J. J. Atick (1992).
%Could information theory provide an ecological theory of sensory processing?,
%%in
%{\it Princeton Lectures on Biophysics,} W. Bialek, ed., pp. 223-289 (World
%Scientific, Singapore).

%\bibitem{atick-redlich90}
%J. J. Atick and A. N. Redlich (1990).
%Towards a theory of early visual processing,
%{\it Neural Comp.}  {\bf 2,} 308-320.

%\bibitem{atick-color}
%J. J. Atick, Z. Li, and A. N. Redlich (1992).
%Understanding retinal color coding from first principles,
%{\it Neural Comp.} {\bf 4,} 559-572.


%\bibitem{exp-tails:atta}
%C. W. van Atta and W. Y. Chen (1970).
%Structure functions of turbulence in the atmospheric boundary layer over the
%ocean, {\it J. Fluid. Mech.} {\bf 44,} 145-159.


%\bibitem{barlow-difflimit}
%H. B. Barlow (1952).
% The size of ommatidia in apposition eyes,
%{\it J. Exp. Biol.} {\bf 29,} 667-674.

%\bibitem{barlow53}
%H. B. Barlow  (1953).    Summation and inhibition in the frog's retina  {\it
%%J.
%Physiol.} {\bf 119,} 69-88.



%\bibitem{barlow61}
%H. B. Barlow (1961).
%Possible principles underlying the transformation of sensory messages,
%in {\it Sensory Communication,} W. Rosenblith, ed., pp. 217-234
%(MIT Press, Cambridge MA).





\bibitem{barlow-ferrier}
H. B. Barlow (1981).
Critical limiting factors in the design of the eye and visual cortex,
{\it Proc. R. Soc. Lond. Ser. B} {\bf 212,} 1-34.


%\bibitem{barlow+levick}
%H. B. Barlow  and W. Levick (1969).
%Three factors limiting the reliable detection of light by the retinal ganglion
%cells of the cat, {\it J. Physiol.} {\bf 200,} 1-24

%\bibitem{benardete}
%E. A. Benardete, E. Kaplan, and B. W. Knight (1992).
%Contrast gain control in the primate retina: P cells are not X-like, some
%M-cells are, {\it Vis. Neurosci.} {\bf 8,} 483-486.



\bibitem{annrevs}
W. Bialek (1987).
Physical limits to sensation and perception,
{\it Ann. Rev. Biophys. Biophys. Chem.}
{\bf 16,} 455-478.



\bibitem{santafe}
W. Bialek (1990).
Theoretical physics meets experimental neurobiology,
in {\it 1989 Lectures in Complex Systems,
SFI Studies in the Sciences of Complexity, Lect. Vol. II,} E. Jen, ed.,  pp.
513-595 (Addison-Wesley, Menlo Park CA).




\bibitem{bialek-plb}
W. Bialek (1992).
Optimal signal processing in the nervous system, in {\it Princeton Lectures on
Biophysics,} W. Bialek, ed., pp. 321-401 (World Scientific, Singapore).


\bibitem{bits+brains}
W. Bialek, M. De Weese, F. Rieke and D. Warland (1993). Bits and brains:
Information flow in the nervous system, {\em Physica A }{\bf 200,} 581-593.

\bibitem{bipolar-bj}
W. Bialek and W. G. Owen (1990).
Temporal filtering in retinal bipolar cells:
Elements of an optimal computation?,
{\it Biophys. J.} {\bf 58,} 1227-1233.




\bibitem{science}
W. Bialek, F. Rieke, R. R. de Ruyter van Steveninck, and D. Warland (1991).
Reading a neural code,
{\it Science} {\bf 252,} 1854-1857.



%\bibitem{brz}
%W. Bialek, D. L. Ruderman, and A. Zee  (1991).
%Optimal sampling of natural images:
%A design principle for the visual system?,
%in {\it Advances in Neural Information Processing 3,}
%R. P. Lippman, J. E. Moody, and D. S. Touretzky, eds., pp. 363-369
%(Morgan Kaufmann, San Mateo CA).

\bibitem{bialek+zee2}
W. Bialek and A. Zee (1987). Statistical mechanics and invariant
perception, {\em Phys. Rev. Lett.} {\bf 58,} 741-744.

\bibitem{bialek+zee}
W. Bialek and A. Zee (1990).
Coding and computation with neural spike trains,
{\it J. Stat. Phys.} {\bf 59,} 103-115.

%\bibitem{bonds91}
%A. B. Bonds (1991).  Temporal dynamics of contrast gain in single cells of the
%cat striate cortex,  {\it Vis. Neurosci.} {\bf 2,} 41-55.

%\bibitem{born+wolf}
%M. Born and E. Wolf (1980).
%{\it Principles of Optics, 6th Ed.} (Pergamon Press, Oxford).

\bibitem{borst+egelhaaf-TINS}
A. Borst and M. Egelhaaf (1989).
Principles of visual motion detection, {\it Trends Neurosci.} {\bf 12,}
297-306.


%\bibitem{brillouin}
%L. Brillouin  (1962).   {\it Science and Information Theory}
%(Academic Press, New York NY).

\bibitem{buchner76}
E. Buchner (1976).
Elementary movement detectors in an insect visual system,
{\it Biol. Cybern.} {\bf 24,} 85-101.

\bibitem{buchner84}
E. Buchner (1984).
Behavioural analysis of spatial vision in insects, in {\it Photoreception and
vision in invertebrates,} M. Ali, ed., pp. 561-622 (Plenum, New York).

%\bibitem{buchsbaum}
%G. Buchsbaum and A.  Gottschalk (1983).
%Trichromacy, opponent colour coding and optimum colour information
%%transmission
%in the retina, {\it Proc. R. Soc. Lond. Ser. B} {\bf 220,} 89-113.



%\bibitem{bullock-rel1}
%T. H. Bullock (1970).
%The reliability of neurons,
%{\it J. Gen. Physiol.} {\bf 55,} 584-656.



%\bibitem{bullock-rel2}
%T. H. Bullock (1976).
%Redundancy and noise in the nervous system:
%Does the model based on unreliable neurons sell nature sort?,
%in {\it Electrobiology of Nerve, Synapse, and Muscle,}
%ed. J. Reuben, D. Purpura, M. V. L. Bennett, and E. Kandel, pp. 179-185
%(Raven Press, New York).



%\bibitem{bees+flowers}
%L. Chittka and R. Menzel (1992).
%The evolutionary adaptation of flower colours and the insect pollinators'
%%colour vision,
%{\it J. Comp. Physiol. A} {\bf 171,} 171-181.

\bibitem{donner-review}
K. Donner (1989). The absolute sensitivity of vision: Can a frog become a
perfect detector of light induced and dark rod events?, {\em Phys. Scr.}
{\bf 39,} 133-140.

%\bibitem{shapley-rev}
%C. Enroth-Cugell and R. M. Shapley (1984).  Visual adaptation and retinal gain
%controls, {\it Prog. Retinal Res.} {\bf 3,} 263-346.


%\bibitem{feynman-lects}
%R. P. Feynman, R. Leighton, and M. Sands (1963).
%{\it The Feynman Lectures on Physics, Volume I.}
%(Addison-Wesley, Reading MA).

\bibitem{feynman+hibbs}
R. P. Feynman and A. R. Hibbs (1965).
{\it Path Integrals and Quantum Mechanics.}
(McGraw Hill, New York).



%\bibitem {field}
%D. Field (1987).    Relations between the statistics of natural
%images and the response properties of cortical cells,
%{\it J. Opt. Soc. Am. A} {\bf 4,} 2379-2394.


\bibitem{franceschini-facets}
N. Franceschini, A. Riehle, and A. le Nestour (1989).
Directionally selective motion detection by insect neurons,
in {\it Facets of Vision,}
R. C. Hardie and D. G. Stavenga, eds.  pp. 360-390
(Springer-Verlag, Berlin).

\bibitem{geman+geman}
S. Geman and D. Geman (1984). Stochastic relaxation, Gibbs distributions,
and the Bayesian restoration of images, {\em I.E.E.E. Trans. P.A.M.I. A}
{\bf 6,} 721-741.

%\bibitem{green+swets}
%D. M. Green and J. A. Swets (1966).
%{\it Signal Detection Theory and Psychophysics.}
%(Wiley, New York).

\bibitem{hass}
S. Hassenstein and W. Reichardt (1956).
Systemtheoretische analyze der zeit-, reihenfolgen-, und vorzeichenaauswertung
bei
der bewegungsperzeption des r\"usselk\"afers {\it Chlorophanus,}
{\it Z. Naturforsch.} {\bf 11b,} 513-524.

\bibitem{hausen+wehrhahn}
K. Hausen and C. Wehrhahn (1983). Microsurgical lesion of horizontal cells
changes optomotor yaw responses in the blowfly {\em Calliphora
erythrocephala}, {\em Proc. R. Soc. Lond. B} {\bf 219,} 211-216.

\bibitem{hausen84}
K. Hausen (1984).
 The lobular complex of the fly:
Structure, function, and significance in behavior,
in {\it Photoreception and vision in invertebrates,}
M. Ali, ed., pp. 523-559 (Plenum, New York).

\bibitem{heisenberg-wolf}
M. Heisenberg and R. Wolf (1984). {\em Vision in {\em Drosophila}:
genetics of microbehavior} (Springer-Verlag, Berlin).

%\bibitem{heslot-exptails}
%F. Heslot, B. Castaing, and A. Libchaber (1987).
%Transitions to turbulence in helium gas, {\it Phys. Rev. A} {\bf 36,}
%%5870-5873.

\bibitem{koch-hildreth}
E. C. Hildreth and C. Koch (1987). The analysis of visual motion---from
computational theory to neuronal mechanisms,
{\em Annu. Rev. Neurosci.} {\bf 10,} 477-533.

\bibitem{horridge-mult}
G. A. Horridge and L. Marcelja (1990). A test for multiplication in
insect directional motion detectors, {\em Phil. Trans. R. Soc. Lond. B}
{\bf 331,} 199-204.

\bibitem{horowitz+hill}
P. Horowitz and W. Hill (1980). {\em The art of electronics,} (Cambridge
University Press, Cambridge UK).

\bibitem{horridge-contrast}
S. Jian and G. A. Horridge (1990). The H1 neuron measures change in velocity
irrespective of contrast frequency, mean velocity or velocity modulation
frequency, {\em Phil. Trans. R. Soc. Lond. B}
{\bf 331,} 205-211.

\bibitem{kac}
M. Kac (1959). {\em Probability and Related Topics in Physical Sciences}
(Interscience, New York).

\bibitem{kolmogorov}
A. Kolmogoroff (1939). Sur l'interpolation et extrapolations des
suites stationnaires, {\em C. R. Acad. Sci. Paris} {\bf 208,} 2043-2045.

\bibitem{kolmogorov2}
A. N. Kolmogorov (1941). Interpolation and extrapolation of stationary
random sequences (in Russian), {\em Izv. Akad. Nauk. SSSR  Ser. Mat.}{
\bf 5,} 3-14. English translation in {\em Selected Works of A. N.
Kolmogorov, Volume II,} A. N. Shiryagev, ed., pp.272-280 (Kluwer Academic
Publishers, Dordrecht, The Netherlands).

%\bibitem{kuffler}
%S. W. Kuffler (1953).   Discharge patterns  and functional organization of
%mammalian retina, {\it J. Neurophys.} {\bf 16,} 37-68.


%\bibitem{kuiper}
%J. W. Kuiper (1962).
%The optics of the compound eye,
%{\it Symp. Soc. Exp. Biol.} {\bf 16,} 58-71.

\bibitem{lackner+zweig}
K. S. Lackner and G. Zweig (1988). Approximating functions from measured
values and prior knowledge, unpublished.

\bibitem{leshouches85}
J.-L. Lacoume, T. S. Durrani and R. Stora, eds. (1985). {\em Traitement
du signal---Signal processing,} Les Houches 1985 Session XLV, Vols. I
and II, (North-Holland, Amsterdam).

%\bibitem{land-handbook}
%M. F. Land (1981).
%Optics and vision in invertebrates,
%{\it Handbk.  Sens. Physiol.} {\bf VII/6B,} 471-592.

%\bibitem{land-in-hbb}
%M. F. Land (1990).  The design of compound eyes,
%in {\it Vision: Coding and efficiency,}
%C. Blakemore, ed., pp. 55-64  (Cambridge University Press, Cambridge UK).



\bibitem{land+collett}
M. F. Land and T. S. Collett (1974).
Chasing behavior of houseflies ({\it Fannia canicularis}):
A description and analysis,
{\it J. Comp. Physiol.} {\bf 89,} 331-357.

%\bibitem{laughlin81}
%S. B. Laughlin (1981).
%A simple coding procedure enhances a neuron's information capacity,
%{\it Z. Naturforsch.} {\bf 36c,} 910-912.

%\bibitem{lawson}
%J. L. Lawson and G. E. Uhlenbeck (1950).
%{\it Threshold Signals.}
%(McGraw-Hill, New York).

\bibitem{lenting-thesis}
B. P. M. Lenting (1985).
{\it Functional characteristics of a wide-field movement processing neuron in
the
blowfly visual system.} (Academisch Proefschrift, Rijksuniversiteit
Gron\-ingen).

%\bibitem{lenting-sat}
%B. P. M. Lenting, H. A. K. Mastebroek, and W. H. Zaagman (1984).
%Saturation in a wide-field, directionally selective movement detection system
%%in
%fly vision, {\it Vision. Res.} {\bf 24,} 1341-1347.


%\bibitem{lettvin}
%J. Y. Lettvin, H. R. Maturana, W. S. McCulluch, and W. H.
%Pitts (1959).     What the frog's eye tells the frog's brain,
%{\it Proc. I. R. E.}  {\bf 47,} 1940-1951.



%\bibitem{linsker89}
%R. Linsker (1989).
%An application of the principle of maximum information preservation to
%linear systems,
%in {\it Advances in Neural Information Processing 1,}
%ed. D. Touretzky, p. 186 (Morgan Kaufmann, San Mateo, CA).
%\marginpar{\bf last pg.}

%\bibitem{ma}
%S.-K. Ma (1976).
%{\it Modern Theory of Critical Phenomena} (Addison-Wesley, Reading CA).

\bibitem{maddess}
T. Maddess and S. B. Laughlin (1985).
Adaptation of the movement-sensitive neuron H1 is generated locally and
governed
by contrast frequency, {\it Proc. R. Soc. Lond. Ser. B} {\bf 225,} 251-275.


%\bibitem{mallock}
%A. Mallock (1894).    Insect sight and the defining power of
%compound eyes,   {\it Proc. R. Soc. Lond. Ser. B} {\bf 55,} 85-90.


%\bibitem{maloney}
%L. T. Maloney (1986).
%Evaluation of linear models of surface reflectance with small numbers of
%parameters, {\it J. Opt. Soc. Am. A} {\bf 3,} 1673-1683.


%\bibitem{ohzawa}
%I. Ohzawa, G. Sclar, and R. D. Freeman (1985).  Contrast gain control in
%the cat's visual system, {\it J. Neurophys.} {\bf 54,} 651-667.

\bibitem{papoulis}
A. Papoulis (1991). {\em Probability, Random Variables, and Stochastic
Processes.}
3rd Ed. (McGraw-Hill)

\bibitem{quartrevII}
T. Poggio and W. Reichardt (1976).
Visual control of orientation behavior in the fly.  Part II.
Towards the underlying neural interactions,
{\it Q. Rev. Biophys.} {\bf 9,} 377-438.


\bibitem{koch+al}
T. Poggio, V. Torre and C. Koch (1985). Computational vision and
regularization theory, {\em Nature} {\bf 317,} 314-319.


\bibitem{quartrevI}
W. Reichardt and T. Poggio (1976).
Visual control of orientation behavior in the fly.  Part I.
A quantitative analysis,
{\it Q. Rev. Biophys.} {\bf 9,} 311-375.


\bibitem{bipolar-nips}
F. Rieke, W. G. Owen, and W. Bialek (1991).
Optimal filtering in the salamander retina,
in {\it Advances in Neural Information Processing 3,}
R. P. Lippman, J. E. Moody, and D. S. Tour\-etzky, eds., pp. 377-383 (Morgan
Kaufmann, San Mateo CA).

\bibitem{fred-thesis}
F. Rieke (1992). {\em Physical principles underlying sensory processing and
computation.} University of California Doctoral Thesis.

\bibitem{europhy}
F. Rieke, D. Warland, and W. Bialek (1993).
Coding efficiency and information rates in sensory neurons,
{\it Europhys. Lett.,} {\bf 22,} 151-156..

\bibitem{optfly}
F. Rieke, D. Warland, R. R. de Ruyter van Steveninck, and W. Bialek (1993).
Optimal processing of visual movement signals:
Theory and experiments in the blowfly,
in preparation.

%\bibitem{taxon}
%K. Rognes and R. E. Blackith (1990).
%{\it Calliphora vicina} Robineau-Desvoidy, 1830 (Insecta, Diptera):
%Proposed conservation of the specific name, {\it Bull. Zool. Nomen.} {\bf 47,}
%187-190.

\bibitem{beyond-nyquist}
D. L. Ruderman and W. Bialek (1992). Seeing beyond the Nyquist limit, {\em
Neural. Comp.} {\bf 4,} 682-690.

\bibitem{scaling}
D. L. Ruderman and W. Bialek (1993).
Statistics of natural images: Scaling in the woods, preprint.

\bibitem{rob-thesis}
R. R. de Ruyter van Steveninck (1986).
{\it Real-time Performance of a Movement-Sensitive Neuron in the Blowfly Visual
System} (Academisch Proefschrift, Rijksuniversiteit Gron\-ingen).

%\bibitem{procroysoc}
%R. de Ruyter van Steveninck and W. Bialek (1988).
%Real-time performance of a movement sensitive neuron
%in the blowfly visual system:
%Coding and information transfer in short spike sequences,
%{\it Proc. R. Soc. Lond. Ser. B}
%{\bf 234,} 379-414.

\bibitem{stepdisc-nips}
R.  R. de Ruyter van Steveninck and W. Bialek  (1992).
Statistical reliability of a blowfly movement-sensitive neuron,  in
{\it Advances in Neural Information Processing 4,}
R. Lippmann, J.  Moody, and D. Touretzky,  eds., pp. 27-34 (Morgan Kaufmann,
San
Mateo CA).

\bibitem{rob-abs84}
R. R. de Ruyter van Steveninck, W. Bialek, and W. H. Zaagman  (1984).
Vernier movement discrimination with three spikes from one neuron, {\it
Perception}
{\bf 13,} A47-48.

\bibitem{rob-H1adapt}
R. R. de Ruyter van Steveninck, W. H. Zaagman, and H. Mastebroek (1986).
Adaptation of transient responses of a movement-sensitive neuron in the visual
system
of the blowfly {\it Calliphora erythrocephela,}
{\it Biol. Cybern.} {\bf 54,} 223--236.


\bibitem{schuling}
F. Schuling, H. Mastebroek, R. Bult, and B. Lenting (1989).
Properties of elementary movement detectors in the fly {\it Calliphora
erythrocephela,}   {\it J. Comp. Physiol. A} {\bf 165,} 179-192.


%\bibitem{shannon}
%C. E. Shannon, (1949).    Communication in the presence of
%noise,   {\it Proc. I. R. E.} {\bf 37,}  10-21.

\bibitem{simmons-nano}
J. A. Simmons, M. Ferragamo, C. F. Moss, S. B. Steveson, and R. A. Altes
(1990). Discrimination of jittered sonar echoes by the echolocating bat,
{\em Eptesicus fuscus}\/: The shape of target images in echolocation,
{\em J. Comp. Physiol. A} {\bf 167,} 589-616.

%\bibitem{snyder77a}
%A. W. Snyder (1977).
%Acuity of compound eyes:  Physical limitations and design,
%{\it J. Comp. Physiol. A} {\bf 116,} 161-182.


%\bibitem{snyder77b}
%A. W. Snyder, D. S. Stavenga, and S. B. Laughlin (1977).
%Information capacity of compound eyes,
%{\it J. Comp. Physiol.} {\bf 116,} 183-207.


%\bibitem{predcod}
%M. V. Srinivasan, S. B. Laughlin, and A. Dubs (1982).
%Predictive coding: A fresh view of inhibition in the retina
%{\it Proc. R. Soc. Lond. Ser. B} {\bf 216,} 427-459.

\bibitem{srini-imp-res}
M. V. Srinivasan (1983)
The impulse response of a movement-detecting neuron and its interpretation
{\em Vision Res.} {\bf 6} 659-663.

%\bibitem{tolhurst}
%D. J. Tolhurst, Y. Tadmor, and T. Chao (1992).
%Amplitude spectra of natural images, {\it Ophthal. Physiol. Opt.} {\bf 12,}
%229-232.

\bibitem{vantrees}
H. L. Van Trees (1968).
{\em Detection, Estimation and Modulation Theory. Part I}
(Wiley, New York)


\bibitem{wagner1}
H. Wagner (1986).
Flight performance and visual control of flight in the free-flying house fly
{\it (Musca
domestica L.).}  I:  Organization of the flight motor, {\it Phil. Trans. R.
Soc. Lond. Ser. B}
{\bf 312,} 527-551.

\bibitem{wagner2}
H. Wagner (1986).
Flight performance and visual control of flight in the free-flying house fly
{\it (Musca
domestica L.).}  II:  Pursuit of targets, {\it Phil. Trans. R. Soc. Lond. Ser.
B}
{\bf 312,} 553-579.

\bibitem{wagner3}
H. Wagner (1986).
Flight performance and visual control of flight in the free-flying house fly
{\it (Musca
domestica L.).}  III:  Interactions between angular movement induced by wide-
and
small-field stimuli, {\it Phil. Trans. R. Soc. Lond. Ser. B} {\bf 312,}
581-595.



%\bibitem{westheimer}
%G. Westheimer  (1981).  Visual hyperacuity, {\it Prog. Sens. Physiol.} {\bf
%%1,}
%1-30.


\bibitem{wiener-time}
N.  Wiener (1949).
{\it Extrapolation, Interpolation and Smoothing of
Time Series.}
(Wiley, New York).

%\bibitem{wilson75}
%K. G. Wilson (1975).
%The renormalization group, critical phenomena, and the Kondo problem, {\it
%Revs. Mod. Phys.} {\bf 47,} 773.

\bibitem{zaagman78}
W. H. Zaagman, H. A. K. Mastebroek, and J. W. Kuiper (1978).
On the correlation model:  Performance of a movement-detecting neural element
in
the fly visual system, {\it Biol. Cybern.} {\bf 31,} 163-168.

\bibitem{zaagman83}
W. H. Zaagman, H. A. K. Mastebroek, and R. R. de Ruyter van Steveninck (1983).
Adaptive strategies in fly vision:  On their image processing qualities, {\it
I.
E. E. E. Trans. Sys. Man Cybern.} {\bf SMC13,} 900-906.

\end{thebibliography}
\end{document}